%% file: Wiki_paper1_2021-06-15-revision_withappendix.tex
\documentclass[12pt]{article}
\usepackage[a4paper,left=2.0cm,right=2.2cm,top=2.5cm,bottom=2.0cm]{geometry}
\usepackage[T1]{fontenc}
\usepackage{natbib}
\usepackage[USenglish,english]{babel}
\usepackage{rotating}
\usepackage{booktabs}
\usepackage{ae,amscd,amsfonts,amsmath,amssymb,bbm,color,eurosym,graphicx,helvet,mathptmx,ntheorem,paralist,psfrag,slashbox,soul,titlesec,times}
\usepackage[shadow,colorinlistoftodos]{todonotes}                

\def\permille{\ensuremath{{}^\text{o}\mkern-5mu/\mkern-3mu_\text{oo}}}

\usepackage{geometry}										
\usepackage[onehalfspacing]{setspace}										
\usepackage[bottom]{footmisc}

\usepackage{xcolor}
\definecolor{hcolor}{RGB}{25,25,97} 						
\usepackage{hyperref}
\hypersetup{pdftitle={Wikipedia Donations}, pdfauthor={Maximilian Linek and Christian Traxler},
	colorlinks=true,linkcolor=hcolor,citecolor=hcolor,urlcolor=blue,
	breaklinks,pdfstartview={FitH}}

\usepackage[labelfont=bf,font={normalsize,normalfont}]{caption}			
\usepackage[font={normalsize,normalfont}]{subcaption}		

\usepackage{float}											
\usepackage{pdflscape}										
\usepackage{afterpage}										
\usepackage{adjustbox}										

\usepackage{booktabs}										
\usepackage{multicol}										
\usepackage{multirow}										

\allowdisplaybreaks											
\begin{document}


\title{Framing and Social Information Nudges at Wikipedia\thanks{This project would not have been possible without the cooperation of Wikimedia Germany. We would like to thank Till Mletzko and Tobias Schumann for their patient support as well as Eugen Dimant, Egon Tripodi and two anonymous referees for their constructive comments. Financial support form the Thyssen Foundation as well as Hertie School's Faculty Activity Fund is gratefully acknowledged. Georg Peter provided excellent research assistance. An earlier version of this manuscript was circulated under the title ``Are Wikipedia Users Conditionally Cooperative? Evidence from Fundraising Trials''.}}

\author{
Maximilian Linek\thanks{Hertie School, Berlin.}
 \and
Christian Traxler$^\dag$\thanks{Max Planck Institute for Research on Collective Goods; CESifo. \href{mailto:traxler@hertie-school.org}{traxler@hertie-school.org}}
}

\date{June 16, 2021}

\maketitle
\thispagestyle{empty}

\begin{abstract}
\singlespacing
\vspace{-5mm}
We analyze a series of trials that randomly assigned Wikipedia users in Germany to different web banners soliciting donations. The trials varied framing or content of social information about how many other users are donating. Framing a given number of donors in a negative way increased donation rates. Variations in the communicated social information had no detectable effects. The findings are consistent with the results from a survey experiment. In line with donations being strategic substitutes, the survey documents that the negative framing lowers beliefs about others' donations. Varying the social information, in contrast, is ineffective in changing average beliefs.

\vspace{5mm}
\end{abstract}

\newpage
\setcounter{page}{1}

\section{Introduction}

The provision of social information -- information about relevant others' behavior -- is a widely used behavioral policy instrument. It has been successfully applied, e.g., to increase payments of tax debt \citep{Hallsworth2017}, to reduce private energy consumption \citep{Allcott2011} or to raise contributions to online communities \citep{Chen-et-al-2010}.
At the same time, there is a growing body of evidence documenting that social information nudges -- similar to other nudges \citep{Sunstein2017} -- often fail to induce behavioral change \citep[e.g.,][]{bicchieri_nudging_2019, Dimant-et-al-2020, Dur2021, Fellner_2013}.

This paper studies the role of social information in the context of online donations to Wikipedia, the world's largest online encyclopedia. Wikipedia, which receives a lot of attention in social science research \citep[e.g.,][]{Greenstein2012, Gallus2017, Munzert2018}, uses web banners to solicit donations. These banners, which may pop-up at the top of a user's browser window, have been controversially discussed in the behavioral science community:\footnote{See, among many others, this \href{https://twitter.com/R_Thaler/status/1074827695282769920}{tweet} by Richard Thaler or this \href{https://evanwarfel.medium.com/wikipedias-appeal-87ec37cb48dd}{blog post} by Evan Warfel.} the banner texts typically use a negative framing and refer to low donation rates. As long as donations are strategic compliments, basic nudge logic suggests that one could generate more donations by employing a positive frame, emphasizing information that points to a higher donation frequency.

Our paper presents the results from a series of trials (A/B tests) run as part of Wikipedia's online fundraising campaigns between 2014 and 2018. We cooperated with Wikimedia Germany (henceforth WMDE), which operates the German language websites of Wikipedia.\footnote{WMDE hosts more than 2.5 million articles, receives more than 1 billion monthly page views, and accounts for almost 8\% of the global Wikipedia traffic (\href{https://de.wikipedia.org/wiki/Wikipedia:Statistik}{Wiki:stats}).} WMDE shared with us data on all trials that systematically varied the framing or the provided social information in their banner texts. In addition, WMDE allowed us to implement a pre-registered trial. In total, we analyze six trials that produced nearly 23,000 donations, summing up to 442,167 Euro.

Two of the trials focused on the framing of a given piece of information, either communicating that `\textit{only}' (negative) or `\textit{already} $N$ people have donated so far' (positive frame). Two further trials communicate (using a negative frame) a given number of donations but vary the reference number of (cumulative versus daily) banner impressions. The treatment banners imply donation-to-impression rates that are between 20 to 40 times higher than the corresponding rates from the control group banners. The fifth trial explains that `less than 0.1\%' (control) or `less than 1\%' (treatment) of users donate. The sixth trial closely follows the spirit of \cite{frey_social_2004} and communicates either a lower (359K) or a higher total number of donors (`more than 400K'; based on the result from last year and the average numbers over the last years, respectively).

There is a long list of behavioral mechanisms suggesting that the interventions could yield positive treatment effects. If the communicated information -- or the positive framing of a given information -- alters beliefs about the conduct of relevant others this could, in principle, increase the perceived strength of conformity pressure \citep{bernheim_theory_1994} or a social norm to donate \citep{bicchieri_2005}. A higher donation rate or more donations might also alter the perceived quality of the provided public good \citep[e.g.,][]{vesterlund_informational_2003}, which could increase an altruistic donor's motivation to donate. These channels would all imply strategic complementarity: providing social information which suggests that relatively many people donate should increase donation rates.

A fundamentally different prediction can be derived from self-signaling models \citep[e.g.,][]{benabou_incentives_2006,benabou_identity_2011}, which might be quite relevant in a context of anonymous online choices.\footnote{There is no perfect anonymity, of course, since donors provide their names and payment information. Nevertheless, there is certainly no direct, human interaction involved and non-donors remain completely anonymous.} If the self-signal return from donating is larger when relatively few others donate (e.g., because users perceive their own donation to be more crucial in supporting WMDE or keeping it independent of commercial advertisement), donations could become strategic substitutes -- and we should observe negative treatment effect.

In addition to these mechanisms, there are numerous theoretical and practical reasons why donations might neither be strategic complements nor substitutes. In our context, the communicated donation rates are very low; it is thus unclear wether the descriptive, empirical information translates into an appropriate normative expectation \citep{bicchieri_right_2009}. It further remains unclear whether social information on the conduct of the general population is credible and relevant; a single user might not feel socially `close' to the overall population \citep{Dimant2019, Goette_2019}. Moreover, the treatments varying the reference number of banner impressions might alter the perceived value of the public good in an unintended direction.
From a more practical perspective, one has to note that Wikipedia's banners include a fairly long text. It is questionable if (and how many) users carefully read and process all the information contained in the banners. Hence, the intention-to-treat (ITT) effects observed in the trials might simply reflect that a minor variation in content is ineffective in altering beliefs and behavior. All these arguments would point to null results.


The data from the trials provide a mixed picture. On the one hand, we find that a negative framing \textit{increases} click and donation rates (relative to banner impressions). The average amount donated remains almost constant; the negative framing thus raises the total amount donated. On the other hand, we do not find any economically or statistically meaningful effects of varying social information. If anything, the data point to a weak, negative impact from treatments that provide information about (relatively) higher donation rates.


To narrow down the interpretation of these findings, we ran a survey experiment. We worked with a sample of WMDE users from Germany (which was representative for the user population in terms of their age distribution) and exposed these users to a subset of banners from the trials. We then elicited users beliefs and perceptions in several dimensions. The survey results, firstly, provide no indication that the treatments alter the perceived public good value or the credibility of the social information. Second, the data clearly indicate that the strength of different motives to donate -- the perceived warm glow of giving as well as the expected social approval in communication with friends -- are positively associated with beliefs about others' donations. In fact, this pattern holds for both, the donation rate among ones friends as well as the overall level of donations in Germany. The communicated social information thus seems relevant. However, the survey documents that most treatments -- with the notable exception of the framing variation -- fail to alter average beliefs about others' behavior. 

A coherent interpretation of the survey results is that minor text changes in one (out of numerous) pieces of information provided in the banners are ineffective in influencing the average readers' beliefs and, subsequently, donation choices. This appears plausible, as most users tend to skim through the dense banner text -- which is not the information they are primarily seeking when they visit Wikipedia.
This interpretation would suggest that simplification might be a crucial ingredient for an effective usage of social information nudges in these banners.

Our study speaks to several strands of research. Firstly, we contribute to the literature on charitable giving \citep{List2011}. Several studies in this field varied social information about donation amounts \citep[e.g.][]{alpizar_anonymity_2008,shang_field_2009}. Communicating a higher donation -- which may also serve as an anchor -- typically triggers a positive effect on the donated amount. These positive intensive margin effects (how much to donate) tend to be negatively associated with extensive margin decisions (to donate or not to donate): pushing up the average donation tends to reduce the number of donors. 

Studies that vary information on the absolute number or the rate of donors are rare. The most well know is \cite{frey_social_2004}, who experiment in a student population, stressing that either 46\% or 64\% of other students have contributed to charitable funds in the past. The authors report a positive but statistically insignificant treatment effect on donation rates. The effect only turns significant after conditioning on students' past donations. A further, marginally significant effect is reported by \cite{Bartke2017}.\footnote{An interesting and closely related study is \citet{martin_how_2008}, who jointly vary cues about donation amounts as well as the number of donors. They report effects on donation rates as well as on donated amounts. Further evidence is discussed in \citet{heldt_conditional_2005}.} We differ from these contributions in that we focus on a series of interventions in the context of online donations. 


Secondly, finding a negative effect of positive framing on donations -- paired with its positive effect on beliefs regarding the total number of donations --- adds to recent evidence on strategic substitutes in collective action problems \cite[e.g.,][]{Cantoni_2019, Hager-et-al-2019}. Note that these findings point in a different direction than the strategic complementarity highlighted in the earlier public goods literature \citep[e.g.,][]{gachter_conditional_2007}. Our data do not allow us to pin down why positive framing leads to less clicks and donations in our context. Understanding this effect as well as the role of contextual difference to studies reporting (precise) null-effects of framing \citep{Dimant-et-al-2020}  is an important task for future research.

Finally, we contribute to the empirical literature on nudges that fail \citep{Sunstein2017}. Numerous studies, for instance, have found a very limited impact of social information provision on tax compliance \citep[][offer a comprehensive meta analysis]{Zareh2019}. Null results are documented in several lab studies \citep[in particular, in the context of anonymous choices; see e.g.,][]{DADDA2017, Capraro-and-Rand-2018, Dimant-et-al-2020}, highlighting the limits of social norm nudging and belief management strategies \citep[see, e.g.,][]{bicchieri_nudging_2019}. Our survey evidence suggests that the results from WMDE's trials could be due to the limited salience of the provided social information. 

In the remainder of the paper, Section \ref{sec_wiki1_data} first discusses the different trials, their implementation and results. Section~\ref{sec_survey} presents the survey experiment and Section \ref{sec_concl} concludes.

\section{Donation Banner Trials}
\label{sec_wiki1_data}

We study Wikipedia users' decision to donate after being exposed to a donation banner. Such a banner may slide down from the top of a user's browser window during WMDE's trialing and annual fundraising period. A click on the banner opens another website that asks users to enter their payment details and to confirm the donation amount. WMDE provided us with data that track banner impressions, banner clicks and information on (completed) donations. Beyond this information Wikipedia stores (quite differently from other players in big tech) basically no data on their users. In particular, there is no information on which Wiki page was visited before donating. We did not obtain consistent micro data on the donation levels (i.e., we observe the total donation volume and can compute average donations). Hence, our analysis will focus on the binary decision to donate (the `extensive margin') rather than the donated amount (`intensive margin'). In addition, we examine the impact of different banners on click rates (banner clicks relative to impressions).


\subsection{Trials and Randomization}
\label{subsec:randomization}
Our analysis covers all randomized trials conducted by WMDE that systematically varied social information, the framing of this information, or information that provides indirect cues about other users' donation frequency. In addition, we designed, pre-registered and implemented our own trial.\footnote{See \href{https://doi.org/10.1257/rct.3543-1.0}{AEA RCT Registry ID 3543} (November, 2018). Due to technical constraints, we were not able to implement the second trial outlined in the registration. Note further that, different from our pre-registration plan, we did not obtain data on `x-clicks' (number of times a banner was closed by users).} All trials tested a baseline banner (``control'') against a variation (``treatment'') that changed the banner text without affecting size or format of the banner. Table \ref{tab_wiki1_overview} provides an overview of the six trials we analyze (see Online Appendix~C for the layout of the banners). The trials, which were conducted between 2014 and 2018, vary in scale, ranging from 1.2 to 6.6 million impressions per trial.\footnote{Given the baseline donation rates, a simple power analysis (with $\alpha = 5$\%, a power of 80\% and assuming i.i.d.~errors) suggests that the minimum detectable effect (MDE) size of the trials ranged between 7 and 13\% of the control group rate of donations to impressions
(trial 1: 11.14\%, 2: 13.02\%, 3: 6.92\%, 4: 7.48\%, 5: 9.37\%, 6: 11.69\%). In absolute terms, the MDE was between 0.07 and 0.29 donations per 1,000 impressions. \label{fn:power}}  In total, the six trials cover nearly 23,000 donations that amount to 442,167 Euro.

\smallskip
\noindent \textbf{Variation in framing.} \,
The first two trials test framing effects in the communication of a given piece of social information. The manipulations only affect one sentence of the banner text that compares numbers of donors and impressions. More specifically, trials 1 and 2 varied the framing of the (given) number of donors, referring to the number in a negative (``\ldots but only $N$ people donated \ldots '') or positive manner (``already $N$ people donated \ldots ''; see Table \ref{tab_wiki1_overview}).

\medskip
\noindent \textbf{Variation in baseline numbers.} \,
Adopting the negative framing from above, both trials 3 and 4 alter the same sentence that communicates a given number of donors relative to a larger or a smaller baseline: the cumulative total number (control) or the average daily number (treatment) of banner impressions in millions. For a fully rational reader of the banner, the same number of donors should appear larger in the treatment conditions: the numbers imply donor-to-impression ratios that are 20 (trial 3) or 40 times (trial 4) larger than those from the control conditions.\footnote{The communicated numbers were truthful (i.e.~WMDE does not engage in deception). The (daily or cumulative) impression numbers are based on banner roll-outs at a given point in time (with substantial variation in daily banner views). Recall further that these trials were conducted in 2014 and 2015.}

\medskip
\noindent \textbf{Direct variation in social information.} \,
The last two trials directly vary social information. Trial 5 communicates the number of worldwide Wikipedia users and notes that ``less than 0.1\%'' (control) or ``less than 1\%'' (treatment) donate. Trial 6, which we designed in cooperation with WMDE, closely follows the spirit of \cite{frey_social_2004} in communicating a lower and a higher number of donors: the control banner notes that 359.000 users donated last year; the treatment highlights an average of ``more than 400.000'' annual donors ``over the last years''. Both pieces of information are truthful.

\medskip
\noindent \textbf{Randomization.} \,
Randomization was conducted online. During a trial, a certain fraction of users is randomly sampled the first time they load a Wikipedia page. These users are then exposed, with equal chances, to either the control or the treatment banner. WMDE provided us with data on the timing of banner impressions (in 15-minute intervals) for several trials. Similar like other studies in a large-$N$ context, we find several statistically significant but quantitatively negligible imbalances. Overall, the data indicate that banner impressions are well balanced over time.\footnote{This point is documented graphically in the Online Appendix. Figure~\ref{fig:balance-hours} illustrates the distribution of control and treatment banner impressions from Trial 6.} The randomization technique provides no reason why this should be any different for, e.g., different Wikipedia content.

Once a banner appears, a user will continue to face this banner and might only donate after repeated exposure.\footnote{The assigned banner type is stored locally, in a cookie. Unless this information is deleted, a user would be confronted with at most 10 banner impressions. Together with the random sampling (which draws a share $p$ of users [with $p\leq 0.1$ and $p \approx 0.01$ in most trials] into a given trial, implying a $p^{2}/2$ chance of being sampled again and assigned to a different banner after deleting cookies and revisiting the webpage again) and the limited time period of a trial, this reduces the chance of exposure to different banners. Nevertheless, there is a small but non-zero risk that some users might have been exposed to different banners of a given trial.}
This means that donation rates cannot be interpreted in terms of donations per user. As we only observe outcomes relative to total impressions, our data do not allow us to cluster standard errors at the user level. Hence, our inference will tend to over-reject the null of no effect. Accounting for this point, our discussion of the trials' results thus focuses on differences that are significant at the 1\%-level.

\begin{landscape}
\bgroup
\def\arraystretch{1.06}
\begin{table}[!ht]
\centering
\caption{Overview of Trials}
\label{tab_wiki1_overview}
\vspace{-1ex}
\footnotesize
\input{Wiki_paper1_Tab1_2021.tex}

\end{table}
\egroup

\end{landscape}

\subsection{Discussion of Interventions}
\label{subsec:discussion}
The type of variation tested in the six trails is quite heterogenous. The first two trials merely vary the framing of information.\footnote{Note that trials 3 and 4 use the negative frame (`only $N$ people donated'), too. In fact, the treatment banner in trial 3 corresponds to the control banner of trial 2. We exploit this feature in Section~\ref{sec_survey}. \label{fn:trial3}}
The second set of trials communicate a given number of donations relative to different baseline numbers: cumulative versus daily banner impressions. The treatments thus vary indirect cues about others' donation frequency -- assuming that users do the implicit arithmetic and compute the implicit donation rates. 
Finally, trials 5 and 6 vary direct information on donation rates and the number of donors, respectively.

Assuming that the interventions influence users' expectations about others' propensity to donation, the treatments could increase donation rate if donations are strategic complements. Strategic complementarity  \citep[or, in the language of the early public goods literature, `conditional cooperation'][]{gachter_conditional_2007}\footnote{We use the term `conditional cooperation' to refer to an empirical regularity (i.e., strategic complementarity in cooperative or pro-social behavior) rather than a specific mechanism or channel (such as indirect reciprocity). \label{fn:definition-cond-coop}} might emerge via numerous different channels:
intrinsic and extrinsic non-pecuniary motives supporting pro-social norms increase with the level of norm compliance \citep[e.g.,][]{bicchieri_2005,Traxler2010}. The treatments could thus strengthen the (perceived) social norm to support WMDE. In a similar way, information about others' donation rates might strengthen conformity pressure \citep[e.g.,][]{bernheim_theory_1994, Goette_2019}.

A further potential driver of strategic complementarity relates to the perceived quality and value of the provided public good \citep{vesterlund_informational_2003,potters_leading-by-example_2007}. One might argue that a higher donation rate (or, equivalently, a higher number of donors) serves as a signal that Wikipedia offers a more valuable public good. In turn, this could raise altruistic users' motivation to donate. Note, however, that some treatments might actually work in the opposite direction. In particular, the control banners of trials 3 and 4, which emphasize a higher baseline number of user, allude to a larger audience. As compared to the treatment banners (which communicates lower user numbers), the text in the control group might actually indicate a more valuable public good. 

From a practical perspective, one has to keep in mind that the banners contained a lot of information (see Online Appendix~C). In fact, the banner text appears to be overly long. It thus remains unclear how many users carefully read and process the information that is varied in the trials. Hence, our outcome data will only capture intention-to-treat (ITT) effects. Even if the communicated information is processed, it might not necessarily be perceived as credible. For instance, the different donation numbers and rates from Trials 5 and 6 (0.1\% vs 1\%; 359K vs exactly 400K) also imply variation in credibility, in particular, if users would update their beliefs in a self-serving manner \citep{Bicchieri_2020}. 
In addition to credibility issues, the social information might not be relevant. On the one hand, a single Wikipedia user might not necessarily feel socially `close' to the overall population of Wikipedia users in Germany \citep{Dimant2019,Goette_2019}. The treatment variation might thus be irrelevant as it is not indicative for the conduct in the relevant reference group. On the other hand, the information might appear incongruous with the implicit normative expectation \citep{bicchieri_right_2009} or too extreme  (i.e. highlighting very low donation rates) in the sense of \citet{croson_limits_2013}.\footnote{Note that the latter argument does not apply to trial 6, which communicated high absolute numbers.} All these arguments would work against finding positive treatment effects.

Finally, one important argument in favor of donations being strategic substitutes is associated with the role of self-signaling \citep[e.g.][]{benabou_incentives_2006,benabou_identity_2011}.  Self-signaling motives might, in fact, play a key role in the context of online choices, as there is little scope for direct observability (and reward or punishment) by relevant others. In the light of our treatment variation, it appears plausible that the self-identity return from donating might be \textit{smaller} if many others are donating -- e.g., if users perceive their own donation to be less important in maintaining Wikipedia or keeping it independent of commercial advertisement. This argument would imply \textit{negative} treatment effects.

To wrap up, there is a long list of channels and mechanism that could render donations to WMDE either strategic complements or substitutes. In the former case, we would expect positive, in the latter case, negative treatment effects. At the same time, there are several arguments for expecting null results. In the remainder of the paper, we let the data speak. Section~\ref{sec_wiki1_results} first presents the results from the trials. Thereafter we present complementary data from a survey experiment. While the identification of a specific mechanism is beyond the scope of this paper, the survey evidence will help to narrow down the room for interpreting the outcome from the trials.



\subsection{Results of Trials}
\label{sec_wiki1_results}

\textbf{Variation in framing.} $\enspace$
Let us first consider the impact of framing.
Relative to the control banners, which communicated the number of donors in a negative frame (``\ldots but only $N$ people donated \ldots ''), the positive frames of the treatment banners (``already $N$ people donated \ldots '') had a \textit{negative} impact on click rates: banner clicks declined significantly from 1.56 to 1.22 clicks per 1,000 impressions in trial~1 ($-$22\%, $p=0.000$); in trial~2, we observe a decline from 2.36 to 2.18 clicks ($-$8\%, $p=0.014$; see Figure \ref{fig_wiki1-clickrate_framing}).

\begin{figure}[!ht]
\centering
\begin{minipage}{0.65\textwidth}
\caption{Variation in framing -- Donation rates}
\label{fig_wiki1_framing}
\begin{subfigure}{0.42\linewidth}
	 \subcaption{Trial 1}
     \includegraphics[width=\linewidth]{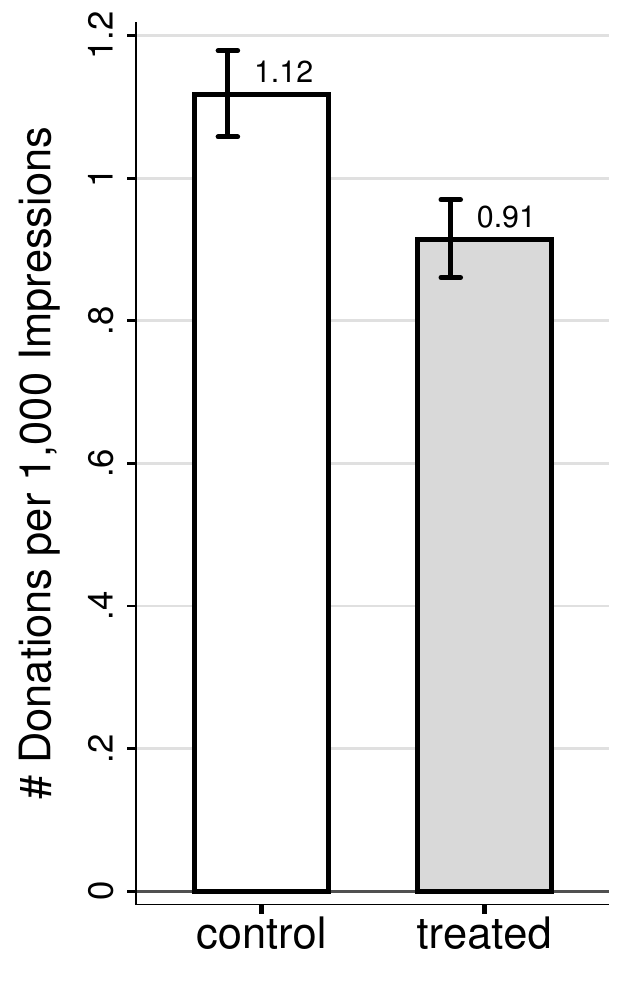}
     \vspace{-4ex}
	 \par\noindent\rule{\textwidth}{0.4pt}
     \scriptsize{Obs.: 2,387,700 \\
				 Chi2: 24.26 ; p-value: 0.000}
\end{subfigure}
\hfill
\begin{subfigure}{0.42\linewidth}
     \subcaption{Trial 2}
     \includegraphics[width=\linewidth]{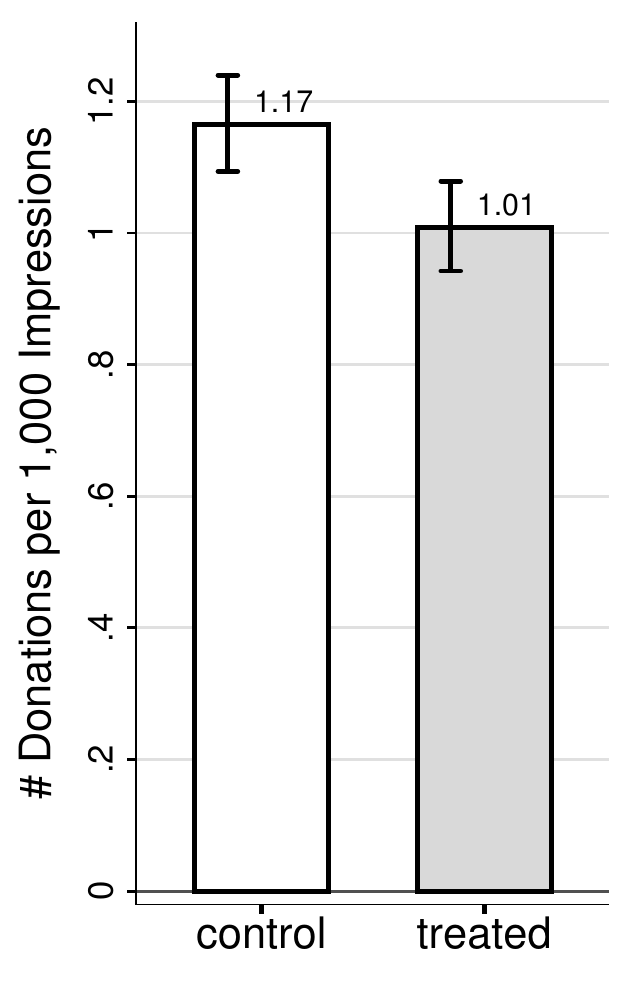}
	 \vspace{-4ex}
	 \par\noindent\rule{\textwidth}{0.4pt}
     \scriptsize{Obs.: 1,691,420 \\
				 Chi2: 9.48 ; p-value: 0.002}
\end{subfigure}
\end{minipage}
\end{figure}

Figure~\ref{fig_wiki1_framing} presents similar results for  donation rates, which dropped from 1.12 to 0.91 donations per 1,000 impressions (–18\%) in trial 1 and from 1.17 to 1.01 donations in trial 2 (–13\%). Keeping the caveat about inference in mind (see Section~\ref{subsec:randomization}), we note that both effects are statistically significant at the 1\%-level. In contrast, the average amount donated is very similar between treatment and control in both trials. The higher amount of total donations in the control conditions reported in Table \ref{tab_wiki1_overview} is thus driven by variation in the extensive rather the intensive margin to donate.

\clearpage

\begin{figure}[!h]
\centering
\begin{minipage}{0.65\textwidth}
\caption{Variation in baseline numbers  -- Donation rates}
\label{fig_wiki1_numbers}
\begin{subfigure}{0.42\linewidth}
	 \subcaption{Trial 3}
     \includegraphics[width=\linewidth]{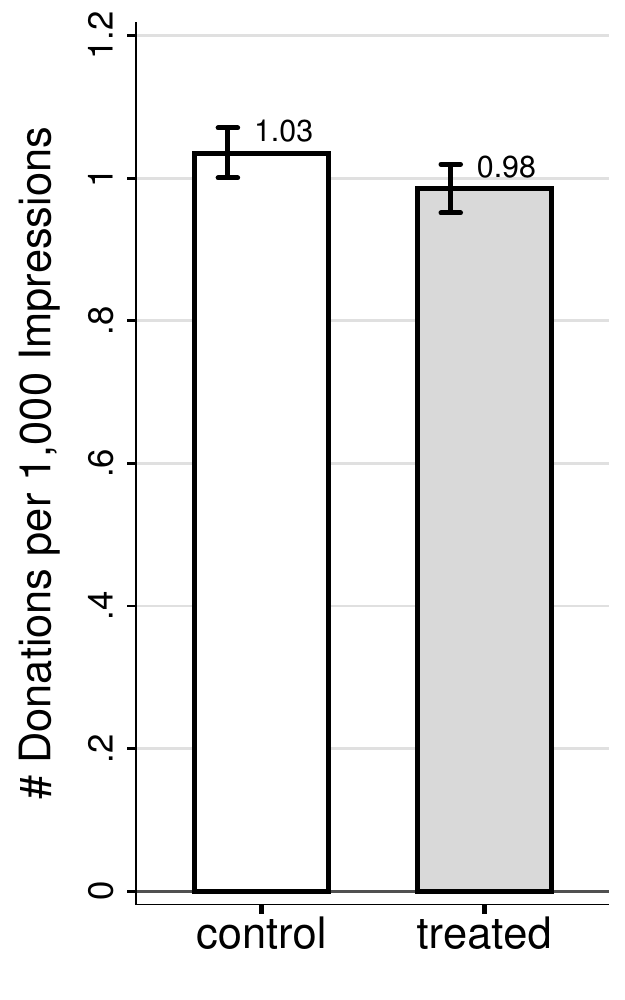}
     \vspace{-4ex}
	 \par\noindent\rule{\textwidth}{0.4pt}
     \scriptsize{Obs.: 6,539,790 \\
				 Chi2: 4.05 ; p-value: 0.044}
\end{subfigure}
\hfill
\begin{subfigure}{0.42\linewidth}
     \subcaption{Trial 4}
     \includegraphics[width=\linewidth]{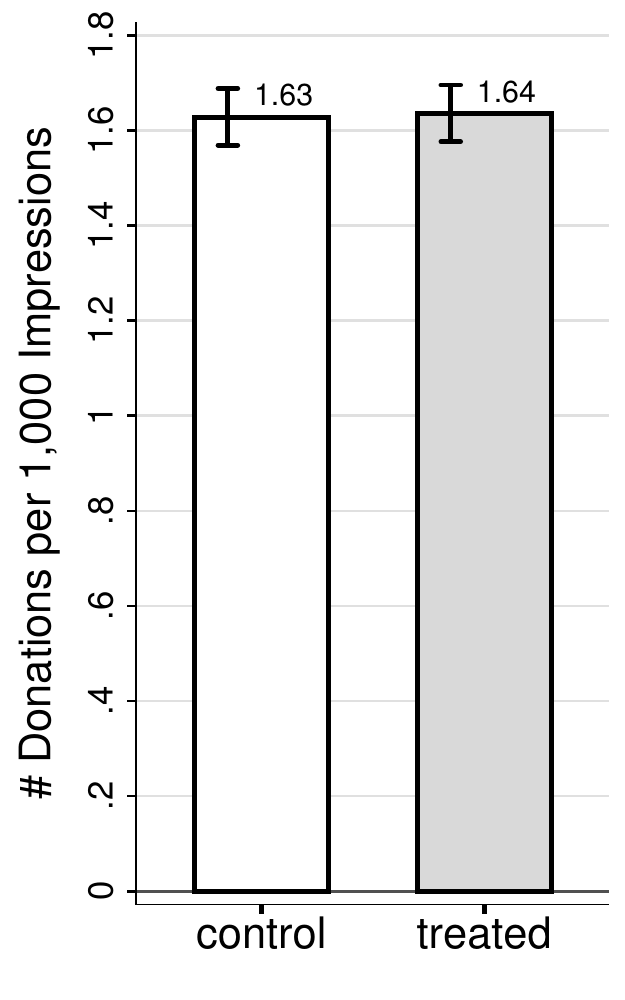}
	 \vspace{-4ex}
	 \par\noindent\rule{\textwidth}{0.4pt}
     \scriptsize{Obs.: 3,565,460 \\
				 Chi2: 0.03 ; p-value: 0.865}
\end{subfigure}
\end{minipage}
\end{figure}


\medskip
\noindent \textbf{Variation in baseline numbers.} $\;$
Next we consider the variation in the communicated number of banner impressions. In terms of click rates, we observe little variation. The data for trial~3 document 2.03 and 1.98 clicks per 1,000 impressions for the control and the treatment banner ($-$2\%), respectively. For trial 4, the corresponding numbers are 2.58 and 2.57 ($-$0.4\%). None of these differences is statistically significant (see Figure \ref{fig_wiki1-clickrate_numbers}).

The results for the donation rates are presented in Figure \ref{fig_wiki1_numbers}. In trial 3, the donation rate dropped from 1.03 to 0.98 per 1,000 impressions (–5\%; $p = 0.044$). This difference reflects a slightly higher conversion rate (of banner clicks into donations) in the control condition.\footnote{The (statistically insignificant) difference in conversion rates can be explained by differential selection of heterogenous types of users into clicking the banner. If extra marginal, treatment responsive users (who are responsive in terms of clicking on the banner) tend to have different conditional probabilities to complete a donation, one obtains variation in conversion rates, despite a common `landing page'.}
In trial 4, we observe a  weak increase from 1.63 to 1.64 (+0.4\%). The average amounts donated and total donations are again similar across the different banners (see Table \ref{tab_wiki1_overview}).



\begin{figure}[!htb]
\centering
\begin{minipage}{0.65\textwidth}
\caption{Direct variation in social information  -- Donation rates}
\label{fig_wiki1_direct}
\begin{subfigure}{0.42\linewidth}
	 \subcaption{Trial 5}
     \includegraphics[width=\linewidth]{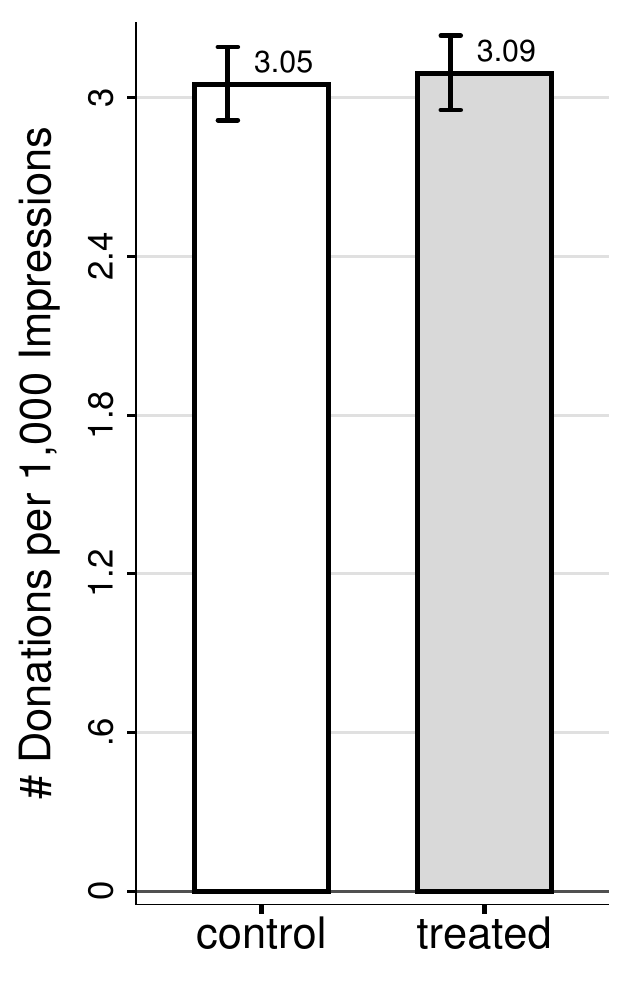}
     \vspace{-4ex}
	 \par\noindent\rule{\textwidth}{0.4pt}
     \scriptsize{Obs.: 1,222,400 \\
				 Chi2: 0.17 ; p-value: 0.684}
\end{subfigure}
\hfill
\begin{subfigure}{0.42\linewidth}
     \subcaption{Trial 6}
     \includegraphics[width=\linewidth]{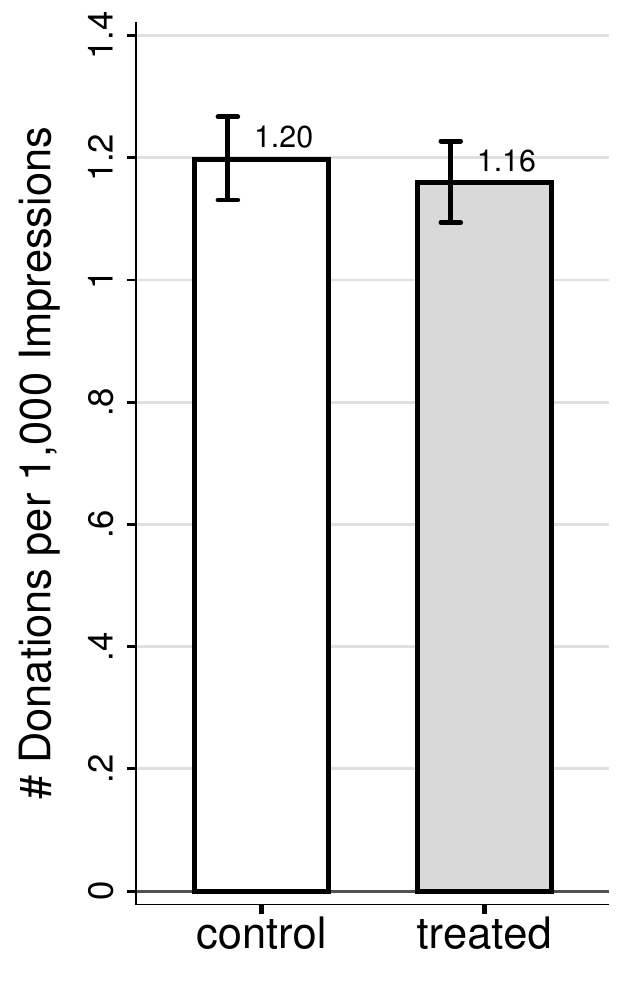}
	 \vspace{-4ex}
	 \par\noindent\rule{\textwidth}{0.4pt}
     \scriptsize{Obs.: 2,026,560 \\
				 Chi2: 0.63 ; p-value: 0.428}
\end{subfigure}
\end{minipage}
\end{figure}


\medskip
\noindent \textbf{Direct variation in social information.}
Trials 5 and 6 directly varied the  social information. 
The former stressed that less than 0.1\% (control) or less than 1\% of users donated (treatment). The latter trial communicated a lower (control) or higher (treatment) absolute number of (yearly) donors in the past. The data on clicks capture again relatively small and statistically insignificant differences. In trial 5, we observe click rates of 4.70 and 4.79 ($+$2\%).\footnote{Click and donation rates in trial 5 are much higher than those observed for the other trials. These differences between trials are hard to interpret. They may be due to different sample periods or variation in banner layouts, colors and size (see Online Appendix~C).} The corresponding rates in trial~6 are 1.63 for the control and 1.64 for the treatment banner ($+$1\%; see Figure~\ref{fig_wiki1-clickrate_direct}).

The results for the donation rates, which are presented in Figure \ref{fig_wiki1_direct}, do not indicate any statistically significant increase either. In trial 5, there are 3.05 donations per 1,000 impressions under the control and 3.09 donations under the treatment banner (+1\%). Finally, in our pre-registered trial, a slightly lower conversion rate in the treatment condition implied a decline from 1.20 donations per 1,000 impressions in the control to 1.16 donations in the treatment group (--3\%). For both trials, the average and total amount donated is once more similar between the different conditions.

\section{Survey Experiment}
\label{sec_survey}
The results from the trials suggest that social information nudges fail to increase banner clicks and donations to WMDE.\footnote{Note the difference to \cite{Dur2021}, who document that a social norm nudge affected clicks but not actual behavioral responses (savings).} In fact, the framing variation from trials 1 and 2 reveals significantly lower click and donation rates under negative frames of a given social information (`only $N$ people donated'). As discussed in Section~\ref{subsec:discussion}, there is a long list of reasons that could explain the null results as well as the negative framing effects. To shed some light on plausible interpretations of the evidence, we ran a survey experiment.

In cooperation with a large online panel provider in Germany, we sampled a population that is representative for Wikipedia users in Germany in terms of their age distribution.\footnote{We used survey data on WMDE usage across different age groups to derive sample quota in the different age bins. In addition, we imposed a balanced gender quota. Individuals indicating that they never used Wikipedia before were excluded from the survey (see Online Appendix~B for details). Let us note that our survey participants faced, similar to Wikipedia users (who visit for the content but then face a donation banner), no incentives to carefully read the full banner text. The external validity of the survey findings to the WMDE user population remains untestable.}
The survey randomly assigned respondents 
to one of 7 banners from the trials discussed in Section~\ref{sec_wiki1_data}. In particular, we implemented the control and treatment banners from trials 2, 5 and 6; we also included the control banner from trial 3. Recall that the corresponding treatment is equivalent to the control banner in trial 2 (see fn.~\ref{fn:trial3}). The survey experiment thus covers the variation from trial 3, too.\footnote{Across all 7 conditions, we used a coherent banner layout (which resembles the one from trials 1, 2, 3 and 5; see Online Appendix C).}

\begin{table}[h!]
\caption{Survey experiment: Overview and match with trials}\vspace{0.75ex}
\label{tab:survey-experiment}
\let\center\empty
\let\endcenter\relax
\centering
\small{
\begin{tabular}{clll}
\toprule
Survey  &                  &  Corresponding  &  \\
Banner $\sharp$ & Obs. & Trial/Banner & Variation  \\
\midrule
1 & 401 & Trial 2 -- Control   & \footnotesize{Neg.~framing ($+$ low baseline number)}\\
2 & 396 & Trial 2 -- Treatment & \footnotesize{Pos.~framing ($+$ low baseline number)}\\
  \hline
3 & 395 & Trial 3 -- Control   & \footnotesize{Neg.~framing $+$ high baseline number }\\
 \hline
4 & 392 & Trial 5 -- Control    & \footnotesize{`Less than 0.1\%' donated}\\
5 & 393 & Trial 5 -- Treatment  & \footnotesize{`Less than 1\%' donated} \\
  \hline
6 & 393 & Trial 6 -- Control    & \footnotesize{`359K people donated' (last year)}\\
7 & 402 & Trial 6 -- Treatment  & \footnotesize{`More than 400K donated' (average over last years)}\\
\bottomrule
\end{tabular}}\\
\vspace{1ex}
\parbox{33pc}{\scriptsize \textit{Notes:} The table lists the banners covered by the survey experiment, the number of observations, as well as the corresponding trials and conditions from the WMDE trials. Note that comparing survey banner $\sharp$1 and $\sharp$3 allows us to examine trial 3.}
\end{table}

An overview of the different banners and their link to the WMDE trials is provided in Table~\ref{tab:survey-experiment}. Balancing checks indicate that randomization was successful (see Table~\ref{tab:survey-sum-stats} in the Online Appendix). The data cover almost 400 observations for each of the 7 banners included in the survey experiment. In total, we recorded 2,772 survey participants, out of which 2,744 completed the survey module (in 5.8min on average; median: 3.9min).\footnote{There were no statistically significant differences in drop-outs between banners.} Exactly 50\% of participants were female, 21\% have an age between 40--49 and almost 13\% are older than 60 (see Table~\ref{tab:survey-sum-stats}).  After exposure to a banner, participants were asked, among others, about their beliefs about donation levels and rates in Germany and among their peers. The survey also elicited perceptions regarding the public good value of the free encyclopedia and regarding participants' responses in case of a donation. (Details are provided in Online Appendix B.2.)

\vspace{-0.3cm}
\subsection{Motives and Beliefs}
Before we study the experimental variation, we first document several patterns in the data. On average, 37\% of respondents agree that donating to WMDE would make them `feel great for doing something good'. This share is positively correlated with respondents' expectations regarding the total number of donors. At the same time, expecting a higher share of friends donating is also correlated with this proxy for warm glow (see Panel a, Table~\ref{tab:survey-correlations} in the Online Appendix).\footnote{The regressions reported in Table~\ref{tab:survey-correlations} pool all survey observations and compare z-scores of the different expectations about others' donations. One of the strongest differences revealed by the data is the one between respondents who indicate a zero vs a non-zero share of friends donating. The former indicate more than 10 percentage point lower rates of warm glow, intentions to communicate and 15 percentage point lower rates of expected social approval among friends (Panel a -- c, Table~\ref{tab:survey-correlations}).}

This first motive for donating does not necessarily require communication between peers. Indeed, only 29\% indicate that banners or donations to WMDE are a topic which is raised in conversations. After donating, however, an average of 27\% would talk to their friends about a donation. Moreover, 34\% would expect to get social approval from their friends (once these friends learn about the donation). The propensity to initiate communication as well as the anticipation of social approval are again positively correlated with the share of friends donating as well as with the expected total number of donors in Germany (see Panel b and c, Table~\ref{tab:survey-correlations}).

The empirical patterns indicate that there clearly is a social component in these potential drivers of donations (warm glow and social approval). In turn, this suggests that there should be scope for social information to influence donation behavior. If a banner would increase beliefs about, e.g., the expected number of donations, this should strengthen non-pecuniary motives to donate.\footnote{Admittedly, this argument relies on correlational evidence. However, when we use the treatment induced variation from Trial 2 and 5 (reported in Table~\ref{tab:survey-effects1-perceptions}) as an instrument, we find an even stronger link between the expected level of donations and the different motives. The correlations reported in Table~\ref{tab:survey-correlations} might therefore represent lower bounds of causal links. (Obviously, this IV approach makes a strong assumption regarding the exclusion restriction, i.e.~no direct effect of the banner text variation on outcomes.)} This leads to the question whether the different banners successfully manipulate beliefs about others' behavior.

\vspace{-0.3cm}
\subsection{Effects on Beliefs}
To examine the  banners' impact on individuals' expectations, we estimate treatment effects using linear probability models following the specification
\begin{equation}
y_i = \alpha + \beta \text{Treatment Banner}_i + \mathbf{X}_i \mathbf{\gamma} + \varepsilon_i,
\label{eq:lpm-treatment}
\end{equation}
where $y_i$ is the dependent variable (typically a measure for beliefs about others' inclination to donate)
and $\mathbf{X}_i$ is a vector of control variables. We estimate this specification separately for each trial (i.e., only including treatment and control of a given trial).\footnote{Pooling data, one obtains very similar results. The presentation and interpretation of these results, however, is more complicated. The same holds for the results from ordered probit models, which again yield similar results.} The coefficient of interest, $\beta$, thus captures the effect of a given $\text{Treatment Banner}_i$ relative to the control group level in a given trial (reflected by the constant term, $\alpha$). The results are reported in Table~\ref{tab:survey-effects1-perceptions}.

The estimates document that the framing variation had a significant effect on the expected total number of donors. The positive framing raises beliefs by around 6\% (see Panel a, Column 5 of Table~\ref{tab:survey-effects1-perceptions}).\footnote{Tables~\ref{tab:survey-effects1-perceptions}, \ref{tab:survey-effects2-motives} and \ref{tab:survey-effects3-PubGood} examine, separately for each trial, treatment effects on 10 outcomes. When we correct for multiple hypothesis testing \citep[using][]{List2019}, the effect remains statistically significant ($p=0.011$).} The estimate hardly changes when we control for observables (Column 6) -- which is consistent with successful randomization.\footnote{This statement applies to all $\beta$ coefficients from Table ~\ref{tab:survey-effects1-perceptions}.} The other belief dimensions (e.g., regarding the share of donors) do not respond to the framing variation (Panel a, Columns 1 -- 4).
For all other tested banner variations (i.e., those from trials 3, 5 and 6) we do not detect any robust effect on beliefs. The estimates for trial 5
point to a weakly significant effect (Panel c, Column 5). Once we account for multiple testing \citep{List2019}, however, the effect is far from conventional significance levels.\footnote{An unrelated but noteworthy observation concerns an inconsistency in stated beliefs: the average survey participant indicates that 8.8\% of their friends donate whereas only 1.4\% of the total user population would donate (compare the coefficients for the constants in Columns 1 and 3, Table~\ref{tab:survey-effects1-perceptions}).}


With the exception of the framing trial, Table~\ref{tab:survey-effects1-perceptions} documents that the social information interventions are ineffective in altering average beliefs. Only for small strata of the data, we do detect some treatment differences. (For instance, if one focuses on the subsample of participants expecting between 300K and 500K donors, 
one observes a visible treatment effect in trial 6; see Figure~\ref{fig:survey-effect-trial6} in the Online Appendix.) Such effects, however, are (i) relatively modest and (ii) they apply only to very small and (iii) (potentially) endogenously selected subsamples. 

Overall, the evidence indicates that the variation in social information failed to influence beliefs regarding others' inclination to donate. The limited impact of the interventions does not seem to be due to a lack of credibility: 52\% of respondents perceive the banner content as (very) credible (only 13\% doubt it) and the different treatments did not significantly affect this assessment (see Columns 7 and 8, Table ~\ref{tab:survey-effects1-perceptions}). A plausible interpretation of the findings is that the average reader, who might be overwhelmed with the lengthy banner text (see Online Appendix C), skims through the text fairly quickly (if at all). The subtle text variations that try to convey different social information might lack sufficient salience to influence beliefs.

The framing variation of a given piece of information had, in contrast, a detectable effect on the overall number of expected donations. 
Recall that the framing variation also had a significant negative impact on click and donation rates in trials 1 and 2. Together with the results from WMDE's trials, the survey evidence therefore supports the notion of strategic substitutes in donations. The findings are inconsistent with strategic complementarity.

\begin{table}[ht!]
\caption{Survey experiment: Treatment effects I}\vspace{0.75ex}
\label{tab:survey-effects1-perceptions}
\let\center\empty
\let\endcenter\relax
\centering
\small{
\begin{tabular}{lcccccccc}
\toprule
Dependent Var.: & \multicolumn{2}{c}{Share, Germany} & \multicolumn{2}{c}{Share, Friends} & \multicolumn{2}{c}{$\log$(Donors)} & \multicolumn{2}{c}{Credibility} \\
                & (1) & (2) & (3) & (4)  &  (5)  &  (6)  &  (7)  &  (8)\\
\addlinespace
\midrule
\multicolumn{4}{l}{\textit{a. Positive Framing -- Trial 2}} \\
\midrule
\addlinespace
Treatment & 0.000 & 0.000 & 0.004 & 0.004 & 0.066*** & 0.064*** & 0.058 & 0.052 \\
          & (0.001) & (0.001) & (0.013) & (0.013) & (0.020) & (0.019) & (0.036) & (0.035) \\
          & [0.921] &         & [0.782] &         & [0.011] &         & [0.633] & \\
\addlinespace
Constant   & 0.014 & 0.014 & 0.089 & 0.089 & 12.46 & 12.46 & 0.479 & 0.482 \\
           & (0.001) & (0.001) & (0.010) & (0.009) & (0.013) & (0.012) & (0.025) & (0.024) \\
\addlinespace
Obs. & 797 & 797 & 794 & 794 & 790 & 790 & 790 & 790 \\
\addlinespace
\midrule
\multicolumn{4}{l}{\textit{b. Higher Baseline Number  -- Trial 3}} \\
\midrule
\addlinespace
Treatment & --0.001 & --0.001 & 0.004 & 0.000 & --0.014 & --0.016 & --0.040 & --0.042 \\
          & (0.001) & (0.001) & (0.013) & (0.013) & (0.019) & (0.018) & (0.036) & (0.035) \\
          & [0.984] &         & [0.960] &         & [0.984] &         & [0.919] & \\
\addlinespace
Constant & 0.015 & 0.015 & 0.086 & 0.088 & 12.48 & 12.48 & 0.519 & 0.520 \\
         & (0.001) & (0.001) & (0.009) & (0.009) & (0.013) & (0.013) & (0.025) & (0.024) \\
\addlinespace
Obs.    & 796 & 796 & 793 & 793 & 790 & 790 & 790 & 790 \\
\addlinespace
\midrule
\multicolumn{4}{l}{\textit{c. Higher Percentage of Donors -- Trial 5}} \\
\midrule
\addlinespace
Treatment & --0.000 & --0.000 & --0.015 & --0.014 & 0.036* & 0.037* & 0.032 & 0.038 \\
          & (0.001) & (0.001) & (0.014) & (0.014) & (0.021) & (0.020) & (0.036) & (0.035) \\
          & [0.993] &         & [0.891] &         & [0.491] &         & [0.912] & \\
\addlinespace
Constant & 0.012 & 0.012 & 0.094 & 0.094 & 12.45 & 12.45 & 0.506 & 0.504 \\
         & (0.001) & (0.001) & (0.011) & (0.010) & (0.014) & (0.014) & (0.026) & (0.025) \\
\addlinespace
Obs.  & 785 & 785 & 782 & 782 & 779 & 779 & 777 & 777 \\
\addlinespace
\midrule
\multicolumn{4}{l}{\textit{d. Higher Number of Donors -- Trial 6}} \\
\midrule
\addlinespace
Treatment & 0.001 & 0.001 & 0.015 & 0.010 & 0.012 & 0.010 & --0.020 & --0.024 \\
          & (0.001) & (0.001) & (0.013) & (0.012) & (0.020) & (0.020) & (0.035) & (0.035) \\
          & [0.852] &         & [0.876] &         & [0.956] &         & [0.936] & \\
\addlinespace
Constant & 0.015 & 0.015 & 0.079 & 0.081 & 12.55 & 12.55 & 0.556 & 0.559 \\
 & (0.001) & (0.001) & (0.009) & (0.009) & (0.014) & (0.014) & (0.025) & (0.025) \\
\addlinespace
Obs. & 795 & 795 & 793 & 793 & 790 & 790 & 789 & 789 \\
\addlinespace
\midrule
Controls        & N & Y & N & Y  & N  & Y  & N  & Y\\
\bottomrule
\end{tabular}}
\\ \vspace{1.5ex}
\parbox{38pc}{\scriptsize \textit{Notes:} The table presents the results from linear probability model estimates of equation \eqref{eq:lpm-treatment}. The dependent variables measure beliefs regarding the share of WMDE users in Germany  (columns 1--2) or the share of friends that donate (columns 3--4), and the (log of the) total number of donors (columns 5--6). Credibility (7--8) is a dummy indicating that the information from the banner is perceived to be credible or very credible. Every second specification includes controls (for gender, age groups and Wikipedia usage categories; see Tab.~\ref{tab:survey-sum-stats}). Robust standard errors are in parentheses. In squared brackets, we report the $p$-values obtained from the multiple hypothesis testing correction proposed by \cite{List2019}. These $p$-values account for the fact that we consider 10 outcome variables (see Tables~\ref{tab:survey-effects1-perceptions}, \ref{tab:survey-effects2-motives}, and \ref{tab:survey-effects3-PubGood}).}
\end{table}

\clearpage

\subsection{Additional Survey Results}
24\% of respondents indicate that it is (very) important to them to donate to WMDE (32\% indicate that it is (very) unimportant). The different treatments did not have any significant effect on this perceived importance (see Table~\ref{tab:survey-effects3-PubGood} in the Online Appendix). Similarly, we do not find any systematic treatment effects on the perceived warm glow from donating or on the propensity to communicate and `harvest' social approval from friends (see Table~\ref{tab:survey-effects2-motives}).\footnote{For trial 5, the data indicate a positive effect on the propensity to communicate (see Columns 3 and 4, Panel c, Table~\ref{tab:survey-effects2-motives}). When we account for multiple testing, however, this effect turns insignificant ($p=0.149$).} Hence, the banner variations do not seem to directly influence these possible motives to donate.

We also examined the perceived value of Wikipedia. On average, 68\% of respondents agree that Wikipedia provides a (very) valuable public good (30\% are neutral and only 2\% consider it worthless). Most banner variations are again ineffective in altering these perceptions. In particular, the treatment banner from trial 3 (which communicates a higher baseline number of users), exerts no positive effect (see Section~\ref{subsec:discussion}).  For trial 5 (which communicates that less than 0.1\% vs.~1\% donate), in contrast, we detect a significant \textit{negative} treatment effect (see Panel c, Table~\ref{tab:survey-effects3-PubGood}). While the observation is hard to interpret, we shall note that it does not translate into an increased importance to donate. 
Finally, 54\% agree that a donation would help Wikipedia to continue to operate without commercial advertisement (less than 12\% disagree). The treatment variation of the different trials had again no effect on this agreement  (see Table~\ref{tab:survey-effects3-PubGood}).

%

\section{Conclusions}
\label{sec_concl}
This paper studied the provision and framing of social information in the context of online donations to Wikipedia in Germany. Our analysis combined data from six trials, which varied the text of donation banners, with data from a survey experiment, that examined the banners' impact on different beliefs and perceptions. The results provide mixed evidence.

One set of trials documents that a positive framing of a given number of donors \textit{reduces} banner clicks and donation rates (as compared to a negative frame). Complementary survey evidence indicates that the positive framing increases beliefs regarding the total number of donors. The results thus suggest that donations might be strategic substitutes rather than complements. Such a result could emerge if the positive framing (`already $N$ users donated') diminishes the self-signaling value of a donation. While the survey data do not corroborate this interpretation, the evidence is nevertheless clear in showing that the negative framing yields a higher overall level of donations.

The other set of trials, which varied different pieces of social information, did not produce any economically or statistically significant effects on click or donation rates. Consistently with these results from the field, the survey evidence documents that the variation in social information had no impact on average beliefs about donation rates among friends or the total user population. Given Wikipedia's lenghty and convoluted banner texts, tiny text variations that aim at communicating different social information seem to lack salience. In order to effectively alter beliefs -- and, eventually, influence donation behavior -- the communicated social information might require a more simple context in the form of a much shorter and more focused banner text. A test of this hypothesis is up to future research. 

\vspace{2cm}

\begin{singlespacing}
\bibliographystyle{ecta}
\bibliography{wiki_cooperation}
\end{singlespacing}

\newpage

\appendix
	\renewcommand\thefigure{\Alph{section}\arabic{figure}}
	\renewcommand\thetable{\Alph{section}\arabic{table}}
\setcounter{page}{1}


\section*{Online Appendix A: Complementary Figures}
\label{sec-appendix-A}
\setcounter{section}{1}
\setcounter{table}{0}
\setcounter{figure}{0}


\begin{figure}[ht!]
\let\center\empty
\let\endcenter\relax
\centering
	  \caption{Impressions by Time Interval and Banner Condition -- Trial 6}\vspace{0.75ex}
	  \label{fig:balance-hours}%
     \includegraphics[width=0.75\textwidth]{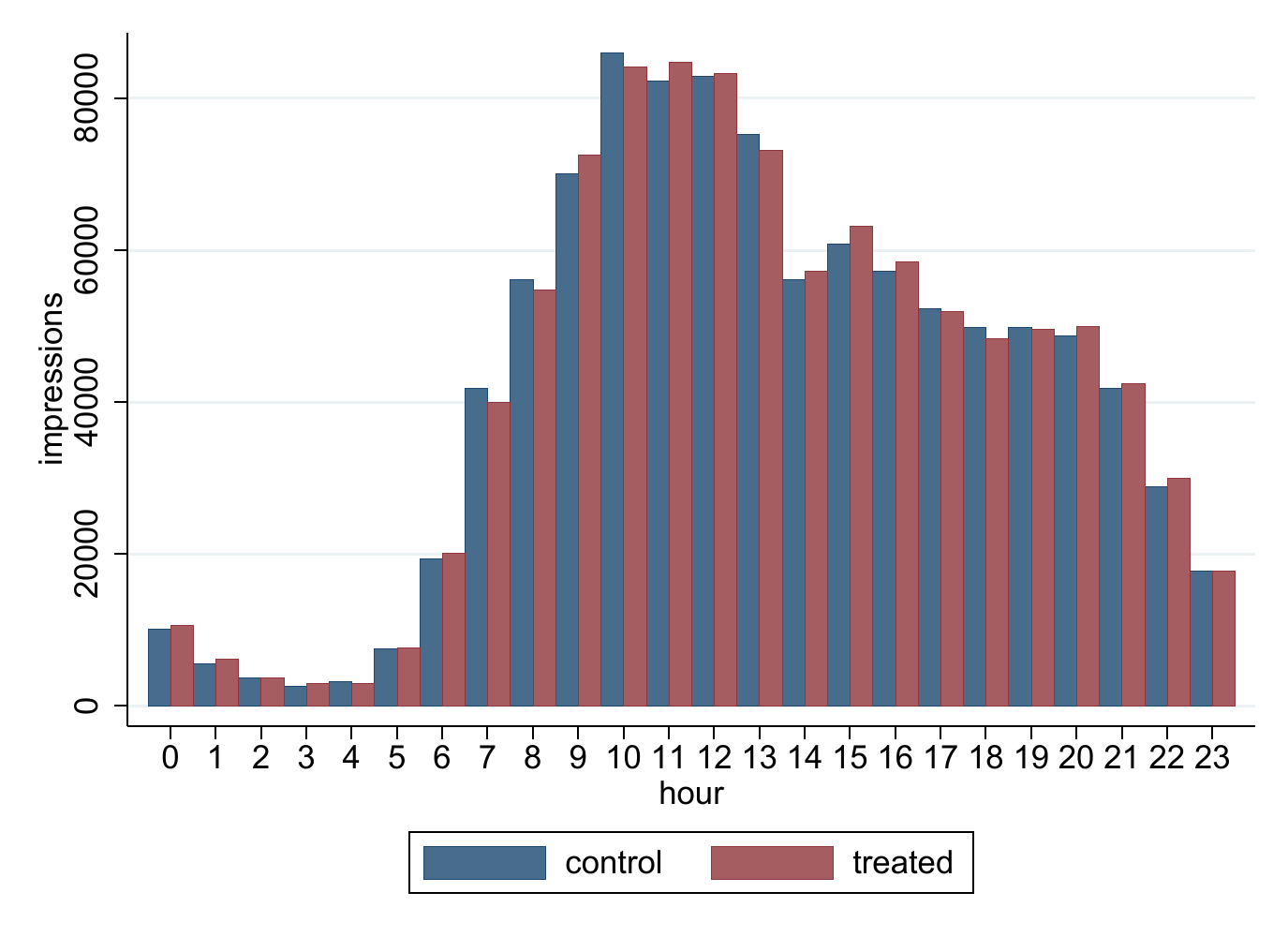}\\
     \includegraphics[width=1.1\textwidth]{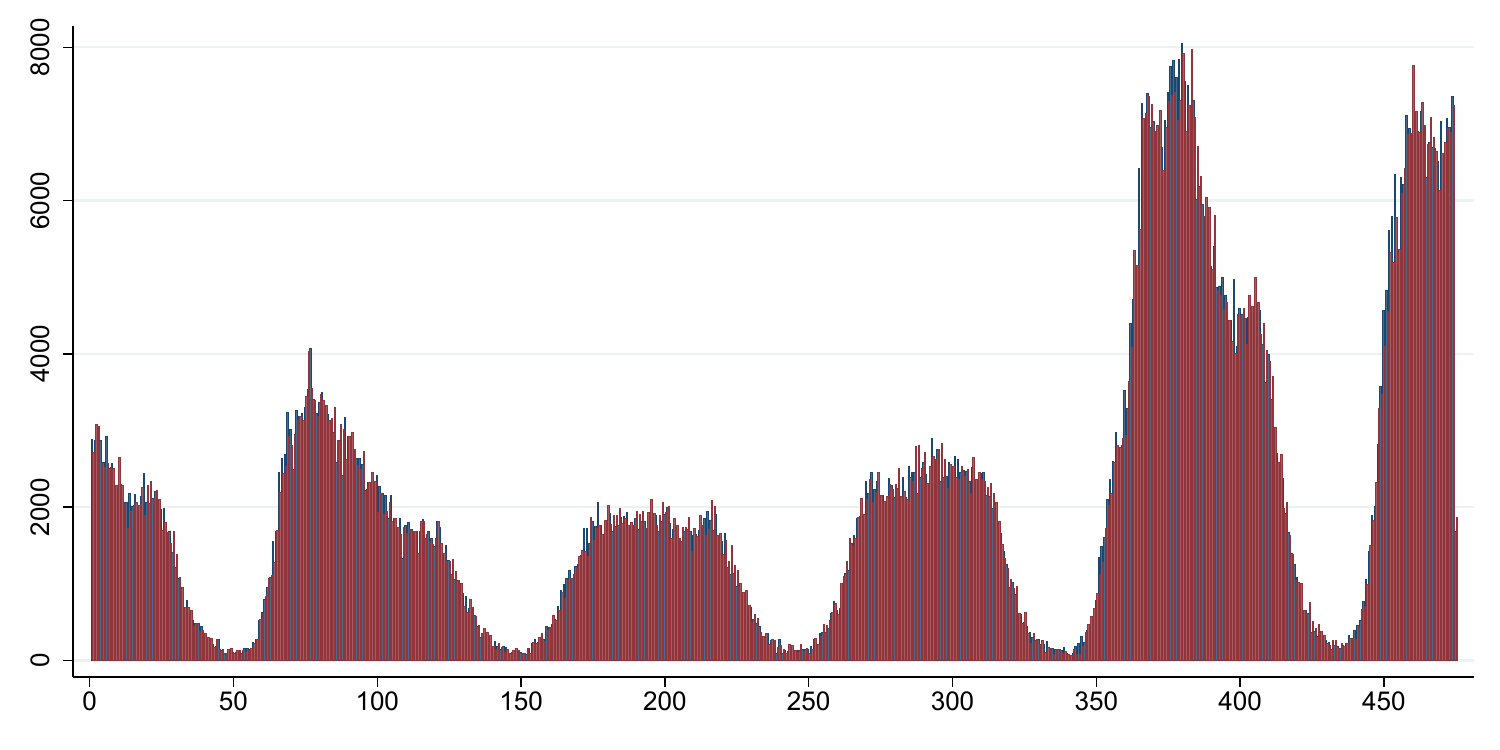}
      \parbox{\linewidth - 1\baselineskip}{\vspace{2ex} \scriptsize
\textit{Notes:} The figures illustrate the number of impressions of Trial 6 for each banner type (treatment vs control) and different time intervals. The top panel depicts impressions for each hour of the day; the lower panel indicates impressions over the entire range of the trial (in 15 minute intervals, the finest resolution we obtained from WMDE). Note that the absence of any major (red or blue) spikes suggests that banner impressions are well balanced over time.}
\end{figure}

\clearpage

\begin{figure}[ht!]
\centering
\begin{minipage}{0.65\textwidth}
\caption{Variation in framing: Effect on click rates}
\label{fig_wiki1-clickrate_framing}
\begin{subfigure}{0.42\linewidth}
	 \subcaption{Trial 1}
     \includegraphics[width=\linewidth]{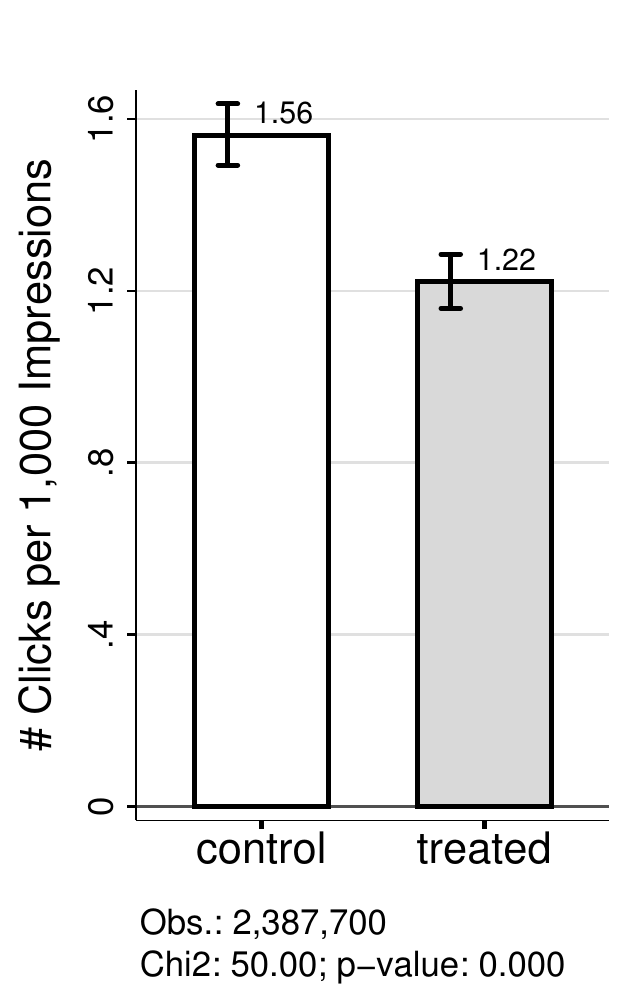}
     \vspace{-4ex}
	 \par\noindent\rule{\textwidth}{0.4pt}
     \scriptsize{Obs.: 2,387,700}
\end{subfigure}
\hfill
\begin{subfigure}{0.42\linewidth}
     \subcaption{Trial 2}
     \includegraphics[width=\linewidth]{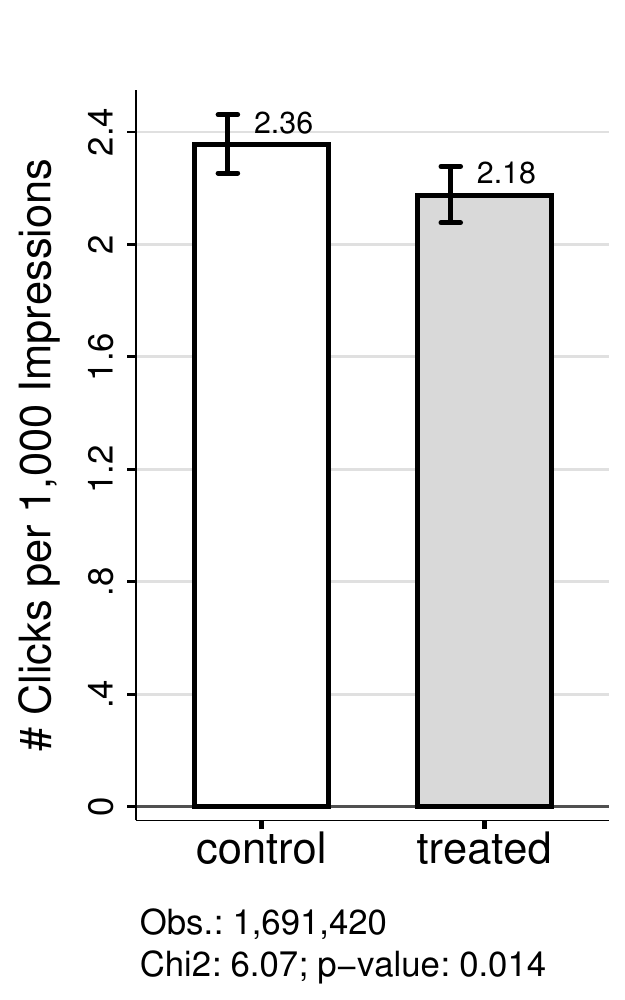}
	 \vspace{-4ex}
	 \par\noindent\rule{\textwidth}{0.4pt}
     \scriptsize{Obs.: 1,691,420}
\end{subfigure}
\end{minipage}
\end{figure}

\begin{figure}[hb!]
\centering
\begin{minipage}{0.65\textwidth}
\caption{Variation in baseline numbers: Effect on click rates}
\label{fig_wiki1-clickrate_numbers}
\begin{subfigure}{0.42\linewidth}
	 \subcaption{Trial 3}
     \includegraphics[width=\linewidth]{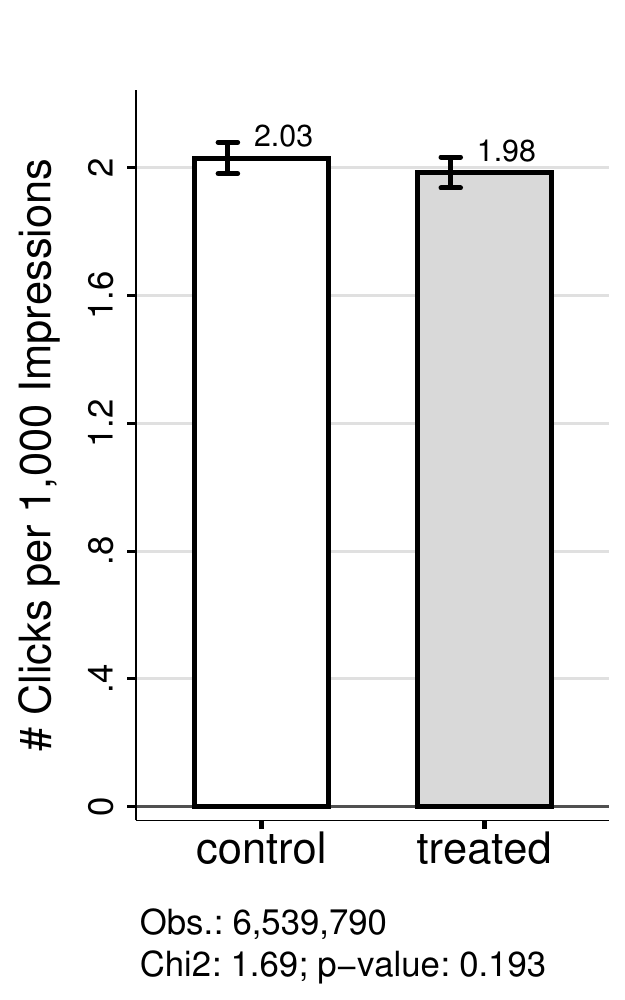}
     \vspace{-4ex}
	 \par\noindent\rule{\textwidth}{0.4pt}
     \scriptsize{Obs.: 6,539,790}
\end{subfigure}
\hfill
\begin{subfigure}{0.42\linewidth}
     \subcaption{Trial 4}
     \includegraphics[width=\linewidth]{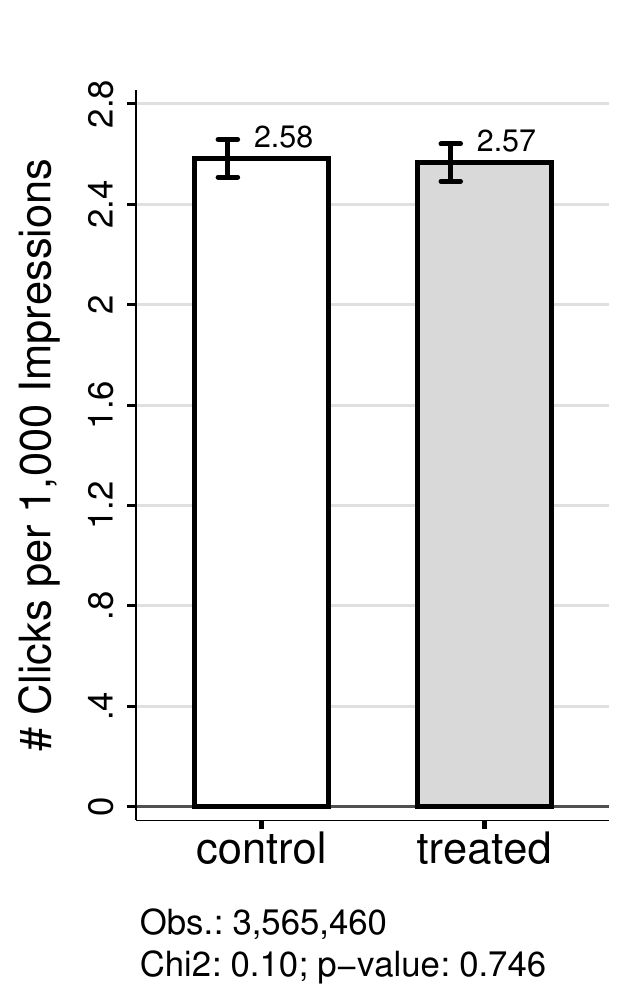}
	 \vspace{-4ex}
	 \par\noindent\rule{\textwidth}{0.4pt}
     \scriptsize{Obs.: 3,565,460}
\end{subfigure}
\end{minipage}
\end{figure}

\clearpage

\begin{figure}[ht!]
\centering
\begin{minipage}{0.65\textwidth}
\caption{Direct variation in social information: Effect on click rates}
\label{fig_wiki1-clickrate_direct}
\begin{subfigure}{0.42\linewidth}
	 \subcaption{Trial 5}
     \includegraphics[width=\linewidth]{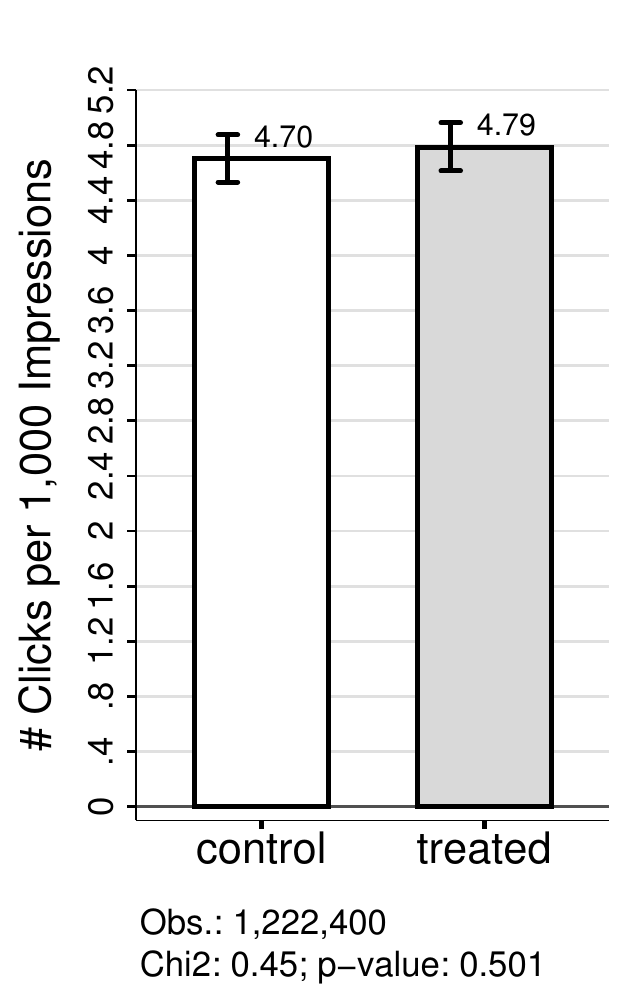}
     \vspace{-4ex}
	 \par\noindent\rule{\textwidth}{0.4pt}
     \scriptsize{Obs.: 1,222,400}
\end{subfigure}
\hfill
\begin{subfigure}{0.42\linewidth}
     \subcaption{Trial 6}
     \includegraphics[width=\linewidth]{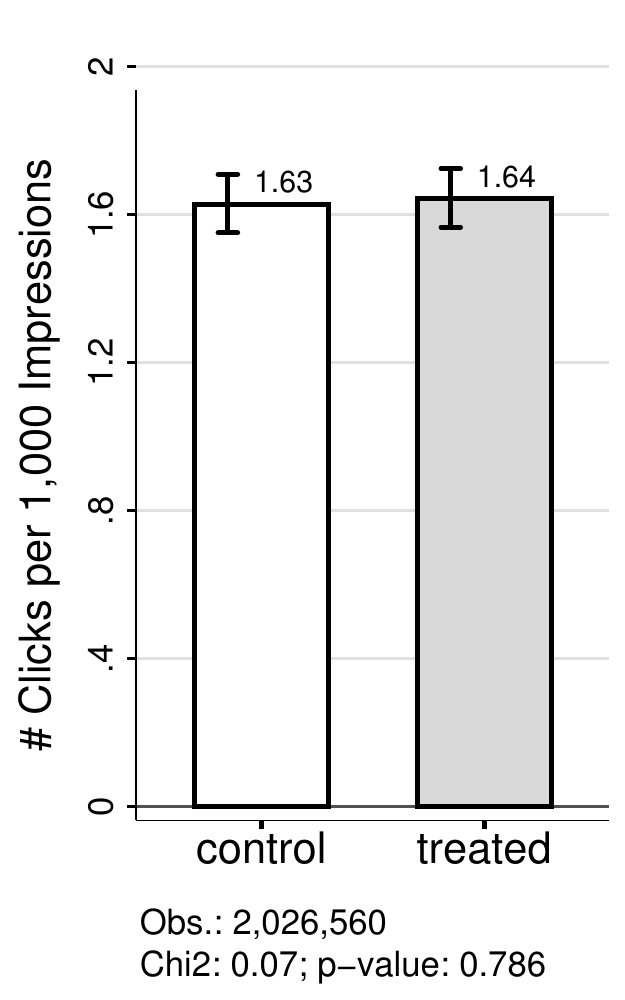}
	 \vspace{-4ex}
	 \par\noindent\rule{\textwidth}{0.4pt}
     \scriptsize{Obs.: 2,026,560}
\end{subfigure}
\end{minipage}
\end{figure}

\clearpage

\newpage

\section*{Online Appendix B: Survey Experiment}
\label{sec-appendix-survey}
\setcounter{section}{2}
\setcounter{table}{0}
\setcounter{figure}{0}

\subsection*{Complementary Figures and Tables: Survey Experiment}
\label{subsec-survey-figs}

\begin{figure}[ht!]
\let\center\empty
\let\endcenter\relax
\centering
	  \caption{Survey experiment: Treatment effect in Subsample -- Trial 6}\vspace{0.75ex}
	  \label{fig:survey-effect-trial6}%
     \includegraphics[width=0.8\textwidth]{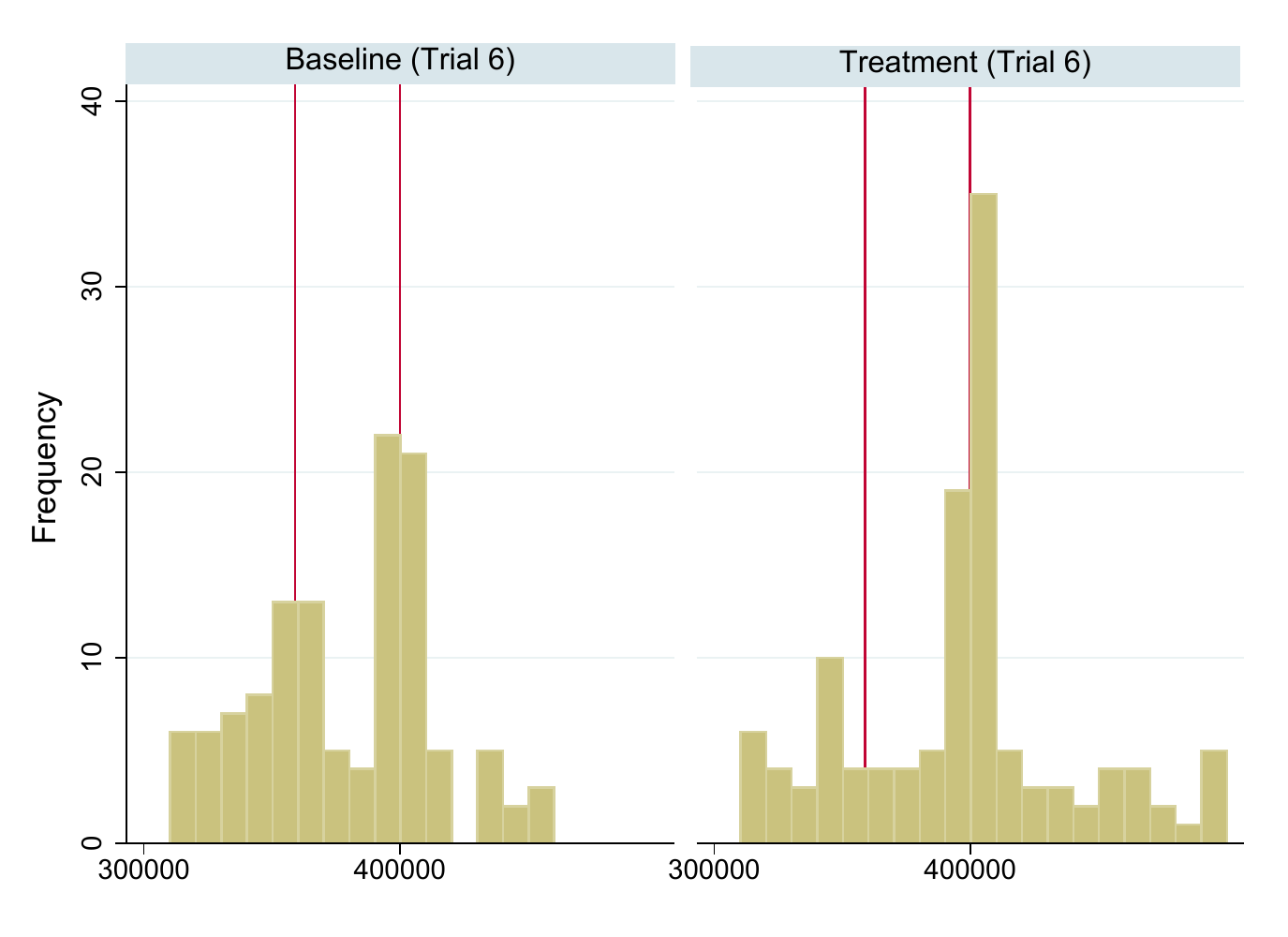}
      \parbox{\linewidth - 1\baselineskip}{\vspace{2ex} \scriptsize
\textit{Notes:} The figure presents responses to the question `How many people in total will donate to Wikipedia in Germany'. The histograms zoom in the subset of responses between 300K and 500K (in bins of 10K). The panels compare the baseline (left) and the treatment banner from trial 6 -- which communicated that 359K (left) and `more than 400K' (right) donated over the last year(s), respectively. While there is clearly some round number heaping in the baseline condition (left panel), there is detectable excess mass around the communicated number (of 359K donations). This mass disappears in the right panel; instead, there is more pronounced heaping at 400K (and also more mass in the range covering `mora than 400K'). Keep in mind that this graph covers only a small and (at least potentially) endogenously selected subset of all responses.}
\end{figure}

\clearpage

\begin{table}[ht!]
\caption{Survey experiment: Summary statistics and balance}\vspace{0.75ex}
\label{tab:survey-sum-stats}
\let\center\empty
\let\endcenter\relax
\centering
\footnotesize{
\begin{tabular}{l|ccccccc|c}
\toprule
Banner $\sharp$: & 1 & 2 & 3 & 4  &  5  &  6  &   7  &  Balance \\
\midrule
female      & 0.489    & 0.490    &  0.496    &  0.526    & 0.501    & 0.519    & 0.480     &  0.436    \\
            & (0.500)  & (0.501)  &  (0.501)  &  (0.500)  & (0.501)  & (0.500)    & (0.500)     &  [0.855]  \\
16--19 years & 0.092    & 0.083    &  0.089    &  0.115    & 0.087    & 0.097    & 0.050     &  2.455     \\
            & (0.290)  & (0.277)  &  (0.285)  &  (0.319)  & (0.281)  & (0.296)    & (0.218)     &  [0.023]   \\
20--29 years & 0.192    & 0.217    &  0.225    &  0.186    & 0.204    & 0.198    & 0.224     &  0.589     \\
            & (0.394)  & (0.413)  &  (0.418)  &  (0.390)  & (0.403)  & (0.399)    & (0.417)     &  [0.739]   \\
30--39 years & 0.185    & 0.162    &  0.200    &  0.202    & 0.165    & 0.191    & 0.221     &  1.154     \\
            & (0.388)  & (0.369)  &  (0.401)  &  (0.402)  & (0.372)  & (0.393)    & (0.416)     &  [0.328]   \\
40--49 years & 0.219    & 0.217    &  0.197    &  0.202    & 0.206    & 0.219    & 0.214     &  0.190     \\
            & (0.414)  & (0.413)  &  (0.399)  &  (0.402)  & (0.405)  & (0.414)    & (0.411)     &  [0.980]   \\
50--59 years & 0.190    & 0.182    &  0.159    &  0.166    & 0.196    & 0.201    & 0.164     &  0.742     \\
            & (0.392)  & (0.386)  &  (0.367)  &  (0.372)  & (0.397)  & (0.401)    & (0.371)     &  [0.615]   \\
$\geq$60 years  & 0.122    & 0.139    &  0.129    &  0.130    & 0.142    & 0.094    & 0.127     &  1.030     \\
            & (0.328)  & (0.346)  &  (0.336)  &  (0.337)  & (0.350)  & (0.292)   & (0.333)     &  [0.404]   \\
Usage-0     & 0.087    & 0.086    &  0.056    &  0.084    & 0.087    & 0.079    & 0.070     &  0.897     \\
            & (0.283)  & (0.281)  &  (0.230)  &  (0.278)  & (0.281)  & (0.270)   & (0.255)     &  [0.496]   \\
Usage-1     & 0.292    & 0.311    &  0.359    &  0.342    & 0.326    & 0.267    & 0.341     &  1.929     \\
            & (0.455)  & (0.463)  &  (0.480)  &  (0.475)  & (0.469)  & (0.443)   & (0.475)     &  [0.073]   \\
Usage-2     & 0.267    & 0.263    &  0.220    &  0.219    & 0.211    & 0.300    & 0.249     &  2.154     \\
            & (0.443)  & (0.441)  &  (0.415)  &  (0.414)  & (0.409)  & (0.459)   & (0.433)     &  [0.045]   \\
Usage-3     & 0.222    & 0.245    &  0.263    &  0.263    & 0.267    & 0.260    & 0.251     &  0.545     \\
            & (0.416)  & (0.431)  &  (0.441)  &  (0.441)  & (0.443)  & (0.439)   & (0.434)     &  [0.774]   \\
Usage-4     & 0.120    & 0.088    &  0.081    &  0.079    & 0.092    & 0.084    & 0.075     &  0.938     \\
            & (0.325)  & (0.284)  &  (0.273)  &  (0.270)  & (0.289)  & (0.278)   & (0.263)     &  [0.466]   \\
Usage-5     & 0.012    & 0.008    &  0.020    &  0.013    & 0.018    & 0.010    & 0.015     &  0.579     \\
            & (0.111)  & (0.087)  &  (0.141)  &  (0.112)  & (0.132)  & (0.100)   & (0.121)     &  [0.748]   \\
\midrule
Obs.        & 401      & 396    &  395     &  392       & 393      & 393        & 402     &          \\
\bottomrule
\end{tabular}}
\vspace{1.5ex}
\parbox{32pc}{\scriptsize \textit{Notes:} The table presents summary statistics (mean and standard deviation in parentheses) for the respondents' gender, age, and Wikipedia usage. The last column presents F-statistic [with p-value in brackets] from separately regressing each of these characteristics on a full set of treatment (banner) dummies, where we test the joint hypothesis that all banner coefficients are equal to zero. Usage categories: `daily' (0), `several times a week' (1),  `at least once a week' (2),  `at least once a month' (3), `at least once a year' (4), `at least once over the last 20 years' (5).}
\end{table}


\clearpage

\begin{table}[ht!]
\caption{Survey experiment: Motives and beliefs about others' donations}\vspace{0.75ex}
\label{tab:survey-correlations}
\let\center\empty
\let\endcenter\relax
\centering
\footnotesize{
\begin{tabular}{lcccccccc}
\toprule
                & (1) & (2) & (3) & (4)  &  (5)  &  (6)  &  (7)  &  (8)\\
\midrule
\addlinespace
\multicolumn{8}{l}{\textit{a. Warm Glow} (`feel great for doing something good'); $N =  2,747$} \\
\addlinespace
\midrule
\addlinespace
Share, Friends$^z$ & 0.043*** & 0.024** &  &  &  &  & 0.019* & 0.003 \\
               & (0.011) & (0.011) &  &  &  &  & (0.011) & (0.011) \\
Share, Germany$^z$ &  &  & 0.036*** & 0.037*** &  &  & 0.004 & 0.017 \\
               &  &  & (0.010) & (0.009) &  &  & (0.011) & (0.011) \\
Number Donors$^z$  &  &  &  &  & 0.074*** & 0.055*** & 0.067*** & 0.047*** \\
               &  &  &  &  & (0.010) & (0.010) & (0.011) & (0.011) \\
Friends Zero & --0.117*** & --0.119*** & --0.130*** & --0.116*** & --0.110*** & --0.108*** & --0.095*** & --0.099*** \\
             & (0.020) & (0.020) & (0.019) & (0.019) & (0.019) & (0.019) & (0.020) & (0.020) \\
Constant & 0.427*** & 0.404*** & 0.433*** & 0.413*** & 0.424*** & 0.394*** & 0.417*** & 0.395*** \\
          & (0.013) & (0.054) & (0.013) & (0.054) & (0.013) & (0.051) & (0.013) & (0.052) \\
\addlinespace
\midrule
\addlinespace
\multicolumn{8}{l}{\textit{b. Tell Friends} (`talk to friends about my donation'); $N =  2,747$} \\
\addlinespace
\midrule
\addlinespace
Share, Friends$^z$ & 0.067*** & 0.056*** &  &  &  &  & 0.048*** & 0.039*** \\
               & (0.010) & (0.011) &  &  &  &  & (0.011) & (0.011) \\
Share, Germany$^z$ &  &  & 0.045*** & 0.044*** &  &  & 0.012 & 0.018* \\
               &  &  & (0.009) & (0.009) &  &  & (0.010) & (0.010) \\
Number Donors$^z$ &  &  &  &  & 0.065*** & 0.052*** & 0.047*** & 0.035*** \\
              &  &  &  &  & (0.009) & (0.009) & (0.010) & (0.010) \\
Friends Zero & --0.086*** & --0.083*** & --0.115*** & --0.101*** & --0.106*** & --0.100*** & --0.069*** & --0.067*** \\
              & (0.018) & (0.018) & (0.017) & (0.017) & (0.017) & (0.017) & (0.018) & (0.018) \\
Constant & 0.305*** & 0.162*** & 0.318*** & 0.185*** & 0.314*** & 0.167*** & 0.297*** & 0.157*** \\
 & (0.012) & (0.039) & (0.012) & (0.039) & (0.012) & (0.039) & (0.012) & (0.039) \\
\addlinespace
\midrule
\addlinespace
\multicolumn{8}{l}{\textit{c. Social Approval} (`friends would approve my donation'); $N =  2,747$} \\
\addlinespace
\midrule
\addlinespace
Share, Friends$^z$ & 0.058*** & 0.038*** &  &  &  &  & 0.037*** & 0.020* \\
               & (0.010) & (0.010) &  &  &  &  & (0.011) & (0.011) \\
Share, Germany$^z$ &  &  & 0.029*** & 0.027*** &  &  & --0.011 & --0.004 \\
               &  &  & (0.010) & (0.009) &  &  & (0.010) & (0.010) \\
Number Donors$^z$ &  &  &  &  & 0.081*** & 0.066*** & 0.074*** & 0.061*** \\
              &  &  &  &  & (0.009) & (0.010) & (0.011) & (0.011) \\
Friends Zero  & --0.152*** & --0.140*** & --0.182*** & --0.154*** & --0.154*** & --0.134*** & --0.133*** & --0.123*** \\
              & (0.019) & (0.019) & (0.019) & (0.018) & (0.018) & (0.018) & (0.020) & (0.019) \\
Constant & 0.407*** & 0.228*** & 0.421*** & 0.243*** & 0.408*** & 0.221*** & 0.399*** & 0.214*** \\
 & (0.013) & (0.046) & (0.013) & (0.046) & (0.013) & (0.048) & (0.013) & (0.047) \\
\addlinespace
\midrule
\addlinespace
Controls        & N & Y & N & Y  & N  & Y  & N  & Y\\
\bottomrule
\end{tabular}}
\\ \vspace{1.5ex}
\parbox{40pc}{\footnotesize \textit{Notes:} The table presents the results from linear probability model estimates with robust standard errors are in parentheses. Every second specification includes controls (for gender, age groups and Wikipedia usage categories; see Tab.~\ref{tab:survey-sum-stats}). The dependent var.~in Panel (a) is a dummy indicating (strong) agreement with feeling `great for doing something good' (in case of a donation); in Panel (b), the dependent var.~indicates (strong) agreement with the statement `I would talk to my friends about my donation'; in Panel (c), the dependent var.~indicates (strong) agreement with friends approving the donation. Variables labeled with $^z$, which measure beliefs about the \textit{Share of Friends}, the \textit{Share of Wikipedia users in Germany} and the total \textit{Number of Donations} expected for Germany, are $z$-normalized ($z$-scores).  \textit{Friends Zero} is a dummy for survey respondents who state that non of their friends would donate to Wikipedia. The number of observations is 2,747 throughout all specifications/panels.}
\end{table}


\clearpage

\begin{table}[ht!]
\caption{Survey experiment: Treatment effects II}\vspace{0.75ex}
\label{tab:survey-effects2-motives}
\let\center\empty
\let\endcenter\relax
\centering
\small{
\begin{tabular}{lcccccc}
\toprule
Dependent Var.: & \multicolumn{2}{c}{Warm Glow} & \multicolumn{2}{c}{Tell Friends} & \multicolumn{2}{c}{Social Approval} \\
  & (1) & (2) & (3) & (4) & (5) & (6) \\
\addlinespace
\midrule
\multicolumn{4}{l}{\textit{a. Positive Framing -- Trial 2}} \\
\midrule
\addlinespace
Treatment  & 0.040 & 0.041 & 0.011 & 0.007 & 0.014 & 0.013 \\
         & (0.035) & (0.034) & (0.031) & (0.031) & (0.033) & (0.033) \\
          & [0.779] &         & [0.979] &         & [0.993] \\
\addlinespace
Constant & 0.363 & 0.362 & 0.249 & 0.251 & 0.307 & 0.308 \\
 & (0.024) & (0.023) & (0.022) & (0.022) & (0.023) & (0.023) \\
\addlinespace
Obs. & 789 & 789 & 789 & 789 & 789 & 789 \\
\addlinespace
\midrule
\multicolumn{4}{l}{\textit{b. Higher Baseline Number  -- Trial 3}} \\
\midrule
\addlinespace
Treatment & --0.022 & --0.027 & --0.033 & --0.030 & --0.016 & --0.023 \\
          & (0.034) & (0.034) & (0.031) & (0.031) & (0.033) & (0.033) \\
          & [0.971] &         & [0.923] &         & [0.976] \\
\addlinespace
Constant & 0.384 & 0.387 & 0.282 & 0.281 & 0.323 & 0.327 \\
 & (0.025) & (0.024) & (0.023) & (0.022) & (0.024) & (0.023) \\
\addlinespace
Obs. & 790 & 790 & 790 & 790 & 790 & 790 \\
\addlinespace
\midrule
\multicolumn{4}{l}{\textit{c. Higher Percentage of Donors -- Trial 5}} \\
\midrule
\addlinespace
Treatment & --0.019 & --0.012 & 0.006 & 0.010 & 0.040 & 0.046 \\
          & (0.034) & (0.033) & (0.031) & (0.031) & (0.034) & (0.032) \\
          & [0.987] &         & [0.985] &         & [0.870] \\
\addlinespace
Constant & 0.358 & 0.355 & 0.249 & 0.247 & 0.312 & 0.309 \\
 & (0.024) & (0.023) & (0.022) & (0.022) & (0.024) & (0.023) \\
\addlinespace
Obs. & 777 & 777 & 777 & 777 & 777 & 777 \\
\addlinespace
\midrule
\multicolumn{4}{l}{\textit{d. Higher Number of Donors -- Trial 6}} \\
\midrule
\addlinespace
Treatment & 0.010 & 0.012 & 0.070** & 0.067** & 0.038 & 0.028 \\
 & (0.035) & (0.034) & (0.032) & (0.032) & (0.035) & (0.034) \\
          & [0.947] &         & [0.149] &         & [0.869] \\
\addlinespace
Constant & 0.379 & 0.378 & 0.246 & 0.248 & 0.356 & 0.362 \\
          & (0.025) & (0.024) & (0.022) & (0.022) & (0.024) & (0.024) \\
\addlinespace
Obs. & 788 & 788 & 788 & 788 & 788 & 788 \\
\addlinespace
\midrule
Controls        & N & Y & N & Y  & N  & Y  \\
\bottomrule
\end{tabular}}
\\ \vspace{1.5ex}
\parbox{32pc}{\scriptsize \textit{Notes:} The table presents the results from linear probability model estimates of equation~\eqref{eq:lpm-treatment}. For the definition of the dependent variables see the notes to Table~\ref{tab:survey-correlations}. Every second specification includes controls (for gender, age groups and Wikipedia usage categories; see Tab.~\ref{tab:survey-sum-stats}). Robust standard errors are in parentheses. In squared brackets, we report the $p$-values obtained from the multiple hypothesis testing correction proposed by \cite{List2019}. The reported $p$-values, which were obtained with the \texttt{mhtexp} package (1000 bootstraps), account for the fact that we consider 10 outcome variables (see Tables~\ref{tab:survey-effects1-perceptions},  \ref{tab:survey-effects2-motives}, and \ref{tab:survey-effects3-PubGood}).}
\end{table}


\clearpage

\begin{table}[ht!]
\caption{Survey experiment: Treatment effects III}\vspace{0.75ex}
\label{tab:survey-effects3-PubGood}
\let\center\empty
\let\endcenter\relax
\centering
\small{
\begin{tabular}{lcccccc}
\toprule
Dependent Var.: & \multicolumn{2}{c}{Importance} & \multicolumn{2}{c}{Public Good} & \multicolumn{2}{c}{Remain Adfree} \\
  & (1) & (2) & (3) & (4) & (5) & (6) \\
\addlinespace
\midrule
\multicolumn{4}{l}{\textit{a. Positive Framing -- Trial 2}} \\
\midrule
\addlinespace
Treatment &  0.040 & 0.035 & 0.040 & 0.031 & 0.025 & 0.023 \\
          & (0.030) & (0.029) & (0.033) & (0.032) & (0.035) & (0.035) \\
          & [0.730] &         & [0.795] &         & [0.971] \\
\addlinespace
Constant   & 0.222 & 0.224 & 0.657 & 0.662 & 0.531 & 0.532 \\
           & (0.021) & (0.020) & (0.024) & (0.023) & (0.025) & (0.025) \\
\addlinespace
Obs. & 790 & 790 & 790 & 790 & 789 & 789 \\
\addlinespace
\midrule
\multicolumn{4}{l}{\textit{b. Higher Baseline Number  -- Trial 3}} \\
\midrule
\addlinespace
Treatment &   --0.010 & --0.020 & --0.035 & --0.036 & 0.002 & 0.002 \\
          & (0.030) & (0.028) & (0.033) & (0.031) & (0.036) & (0.035) \\
          & [0.988] &         & [0.921] &         & [0.936] \\
\addlinespace
Constant & 0.232 & 0.237 & 0.692 & 0.693 & 0.529 & 0.529 \\
 & (0.021) & (0.020) & (0.023) & (0.022) & (0.025) & (0.024) \\
\addlinespace
Obs. & 790 & 790 & 790 & 790 & 790 & 790 \\
\addlinespace
\midrule
\multicolumn{4}{l}{\textit{c. Higher Percentage of Donors -- Trial 5}} \\
\midrule
\addlinespace
Treatment  & 0.004 & 0.006 & --0.097*** & --0.086*** & --0.012 & --0.003 \\
         & (0.031) & (0.029) & (0.034) & (0.032) & (0.036) & (0.035) \\
          & [0.887] &         & [0.075] &         & [0.985] \\
\addlinespace
Constant & 0.251 & 0.250 & 0.712 & 0.706 & 0.540 & 0.536 \\
 & (0.022) & (0.021) & (0.023) & (0.022) & (0.025) & (0.025) \\
\addlinespace
Obs. & 778 & 778 & 777 & 777 & 777 & 777 \\
\addlinespace
\midrule
\multicolumn{4}{l}{\textit{d. Higher Number of Donors ---- Trial 6}} \\
\midrule
\addlinespace
Treatment & 0.024 & 0.009 & --0.025 & --0.026 & --0.004 & 0.002 \\
 & (0.031) & (0.030) & (0.032) & (0.032) & (0.035) & (0.036) \\
          & [0.954] &         & [0.960] &         & [0.910] \\
\addlinespace
Constant & 0.254 & 0.261 & 0.718 & 0.718 & 0.559 & 0.556 \\
 & (0.022) & (0.022) & (0.023) & (0.023) & (0.025) & (0.025) \\
\addlinespace
Obs. & 790 & 790 & 790 & 790 & 788 & 788 \\
\addlinespace
\midrule
Controls        & N & Y & N & Y  & N  & Y  \\
\bottomrule
\end{tabular}}
\\ \vspace{1.5ex}
\parbox{34pc}{\scriptsize \textit{Notes:} The table presents the results from linear probability model estimates of equation~\eqref{eq:lpm-treatment}. The dependent var.~are all dummy variables which indicate: respondents thinking it is (very) \textit{Important} to them to support WMDE with a donation (columns 1--2); respondents noting that Wikipedia provides a (very) valuable \textit{Public Good} (columns 3--4); respondents who (strongly) agree that a donation would help Wikipedia to  \textit{Remain Adfree} (columns 5--6). Every second specification includes controls (for gender, age groups and Wikipedia usage categories; see Tab.~\ref{tab:survey-sum-stats}). Robust standard errors are in parentheses. In squared brackets, we report the $p$-values obtained from the multiple hypothesis testing correction proposed by \cite{List2019}. The reported $p$-values, which were obtained with the \texttt{mhtexp} package (1000 bootstraps), account for the fact that we consider 10 outcome variables (see Tables~\ref{tab:survey-effects1-perceptions},  \ref{tab:survey-effects2-motives}, and \ref{tab:survey-effects3-PubGood}).}
\end{table}


\clearpage

\subsection*{Survey experiment: Details}
\label{subsec-survey-qs}

\begin{enumerate}
\item How frequently do you use Wikipedia?
    \begin{itemize}
      \item {\small{7 response options ranging from `every day' to `never'}}
      \item {\small{Respondents indicating `never' ($<$5\% of total survey clicks) were not admitted to the survey.}}
    \end{itemize}\smallskip
\item How old are you? What is your gender?
    \begin{itemize}
      \item {\small{6 age categories: $<$20 / 20--29 / 30--39 / ... / $\geq 60$ years }}
      \item {\small{To match the age and gender distribution of Wikipedia users in Germany, we imposed quota based on these variables.}}
    \end{itemize}\bigskip

\item[] \textbf{Randomization}
    \begin{itemize}
      \item At this point, participants were randomly assigned to one of 7 banners (see Table~\ref{tab:survey-experiment}).
    \end{itemize}\smallskip

\item {\small{Wikipedia uses banners like the one displayed above to prompt visitors of their webpage to donate.}}\\
      What do you think is the share of Wiki users \textbf{in Germany} that donate to Wikipedia?
    \begin{itemize}
      \item {\small{Response via slider (indicating, e.g., `1 out of 1000 users'; range: 0 -- 50 $\permille$)}}
    \end{itemize}\smallskip

\item What is the share of Wiki users \textbf{among your peers} that donate to Wikipedia?
    \begin{itemize}
      \item {\small{Response via slider (indicating, e.g., `1 out of 100'; range: 0 -- 100\%)}}
    \end{itemize}\smallskip

\item How many people in total will donate to Wikipedia in Germany in 2021?
    \begin{itemize}
      \item {\small{Response via slider (range: 200K -- 600K)}}
    \end{itemize}\smallskip

\item How important is it \underline{to you personally}, to support Wikipedia with a donation?
    \begin{itemize}
      \item  {\small{5 response options (from `very important' to `not at all important') }}
    \end{itemize}\smallskip

\item How valuable is Wikipedia for the \underline{general public}?
    \begin{itemize}
      \item  {\small{5 response options (from `very valuable' to `not at all valuable') }}
    \end{itemize}\smallskip

\item The banner contains several pieces of information. How credible do you think this information is?
    \begin{itemize}
      \item  {\small{5 response options (from `very credible' to `not at all credible') }}
    \end{itemize}\smallskip

\item {\small{Assume you would respond to the banner and donate. Do you agree with the following statements?}}\smallskip
    \begin{itemize}
\item[(a)] I would feel great for doing something good.
\item[(b)] I would talk to my friends about my donation.
\item[(c)] My friends would approve my donation.
\item[(d)] My donation facilitates that Wikipedia can continue to operate without commercial advertisement.\medskip

      \item  {\small{5 response options (from `strongly agree' to `strongly disagree')}}
    \end{itemize}\smallskip

\item Has the topic Wikipedia banners/donations ever been raised by your peers?
    \begin{itemize}
      \item  {\small{5 response options (from `yes, frequently' to `no, never')}}
    \end{itemize}

\end{enumerate}

\newpage

\section*{Online Appendix (C): Banner Layout and Texts}
\label{sec-appendix-C}
\setcounter{section}{3}

The figures below indicate the different banners used in the six trials. All banners of trials 1 to 4 included sliders, which reported (daily adopting) numbers regarding the cumulative donation volume relative to a fixed, pre-specified target value. Except for trial 3, these sliders would also indicate a count to the projected end of the donation campaign (i.e., when the targeted donation total is reached).

\begin{figure}[!ht]
\centering
\caption{Trial 1 -- Control}
\includegraphics[width=\linewidth]{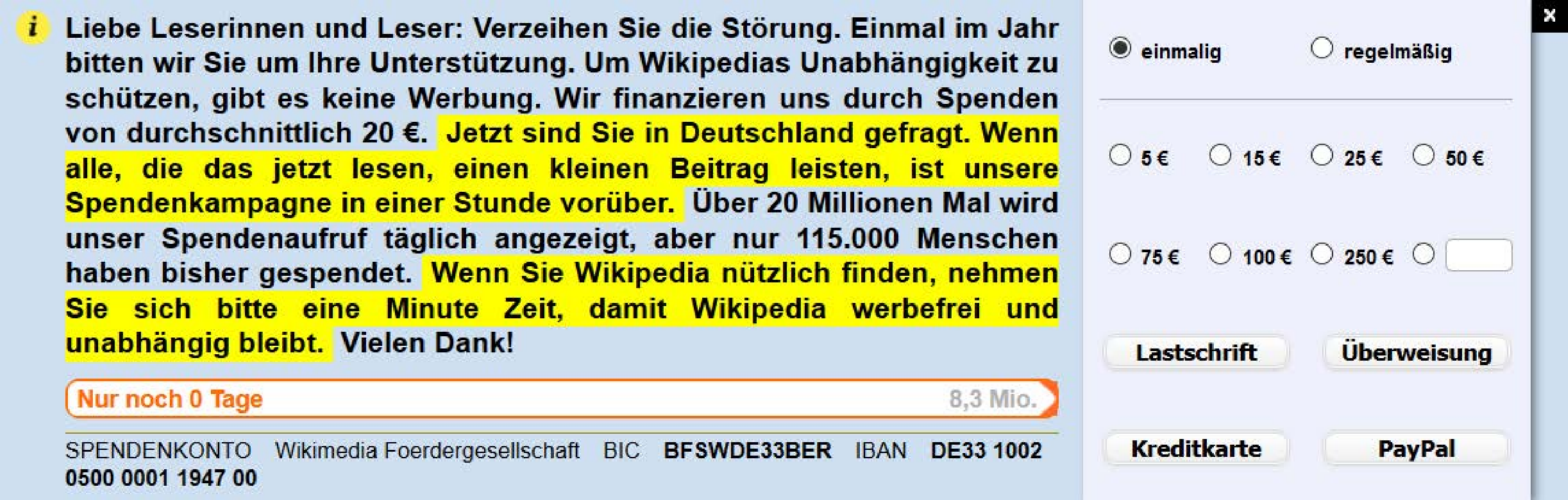}
\end{figure}

\begin{figure}[!ht]
\centering
\caption{Trial 1 -- Treatment}
\includegraphics[width=\linewidth]{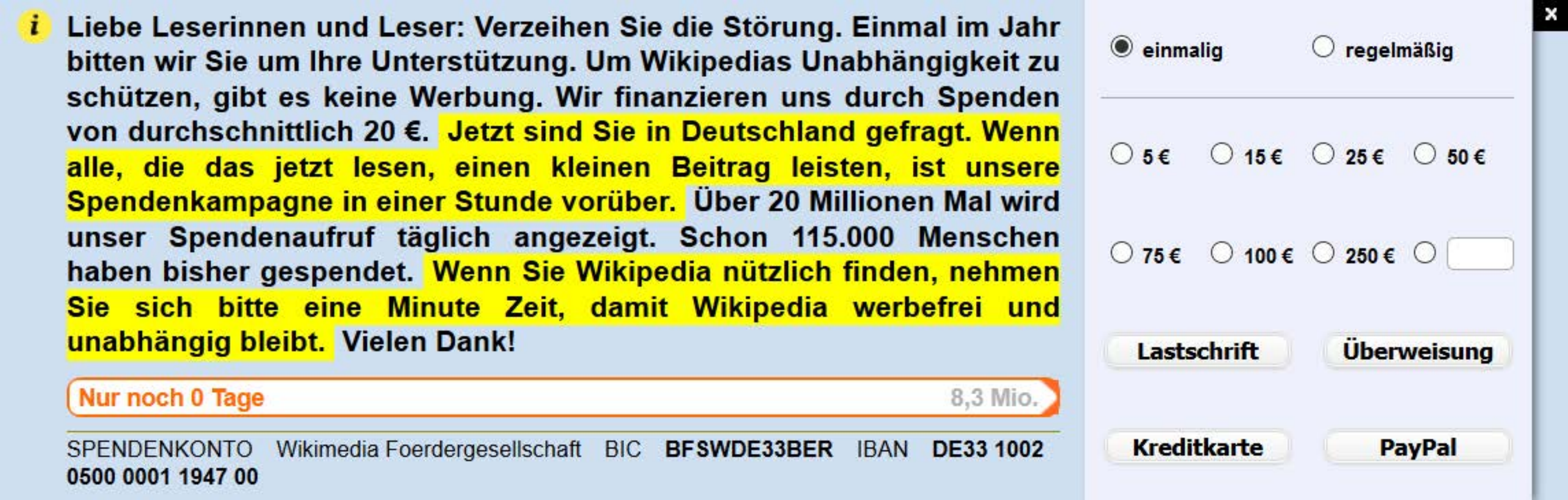}
\end{figure}

\clearpage

\begin{figure}[!ht]
\centering
\caption{Trial 2 -- Control}
\includegraphics[width=\linewidth]{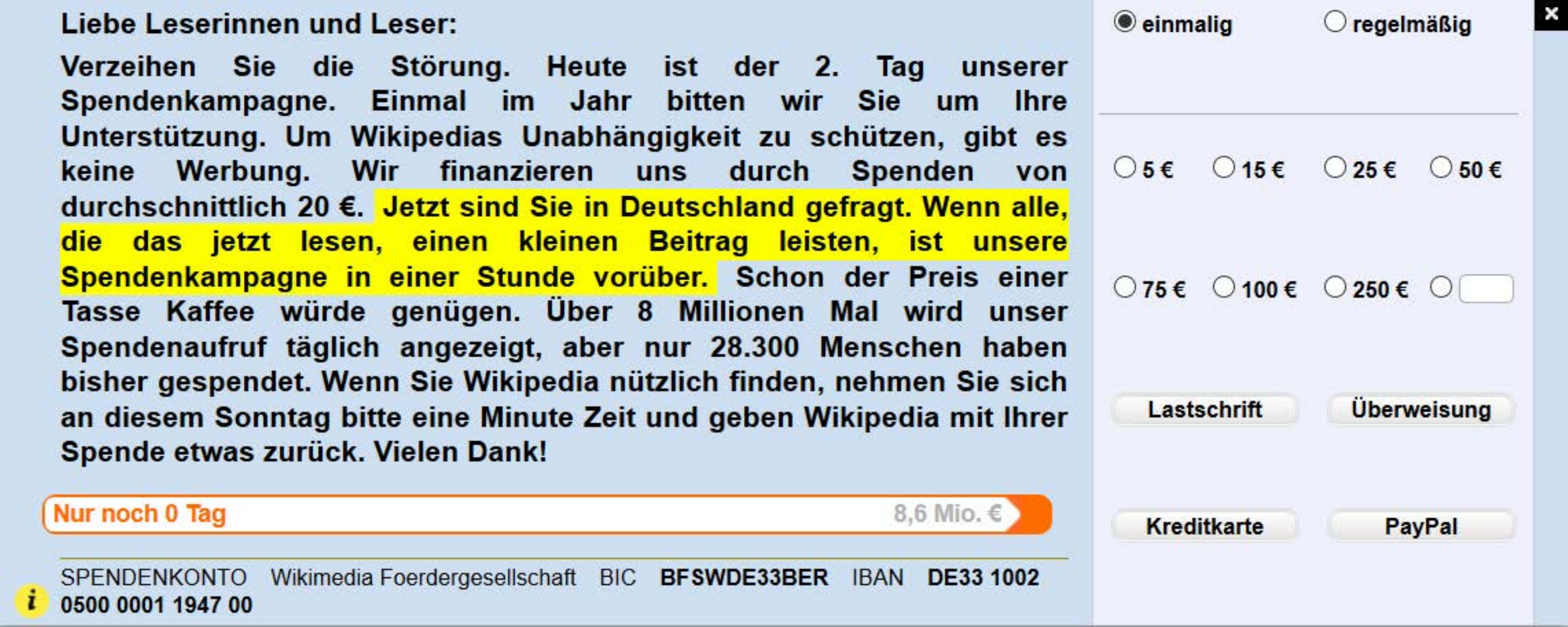}
\end{figure}

\begin{figure}[!ht]
\centering
\caption{Trial 2 -- Treatment}
\includegraphics[width=\linewidth]{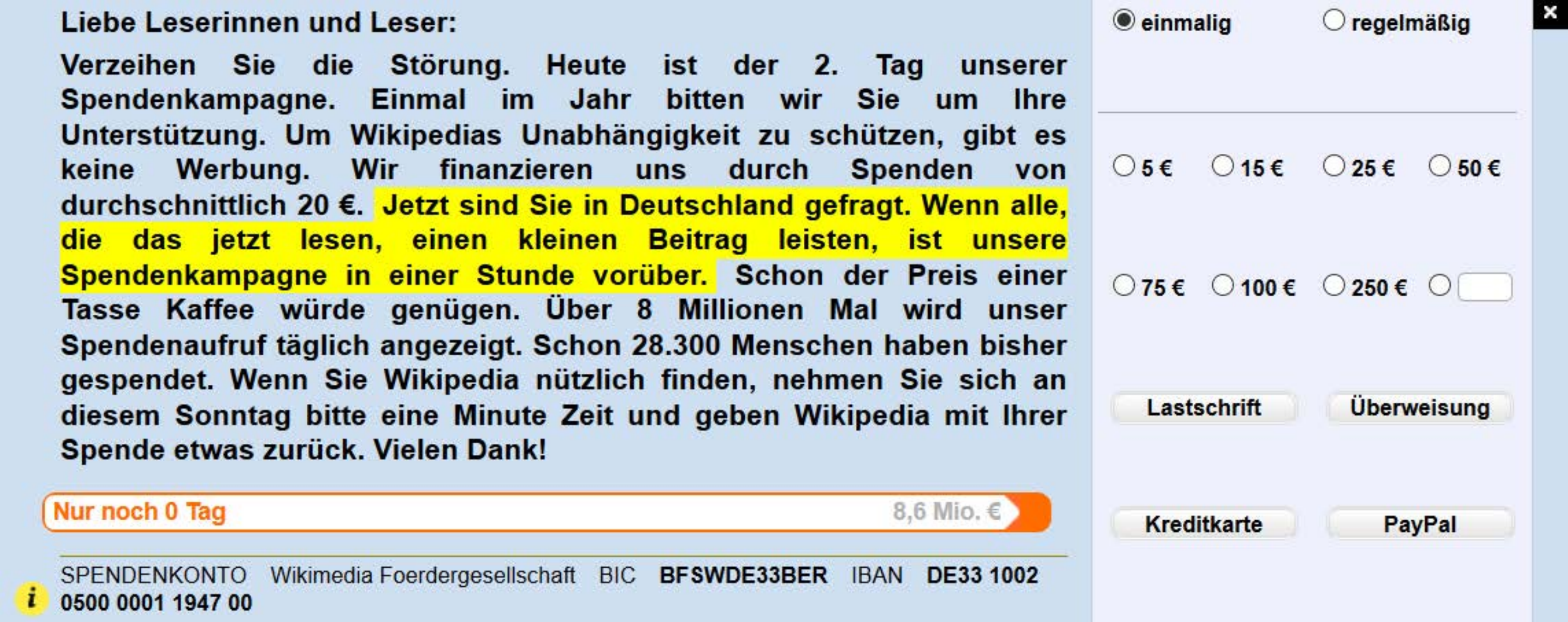}
\end{figure}

\clearpage

\begin{figure}[!ht]
\centering
\caption{Trial 3 -- Control}
\includegraphics[width=\linewidth]{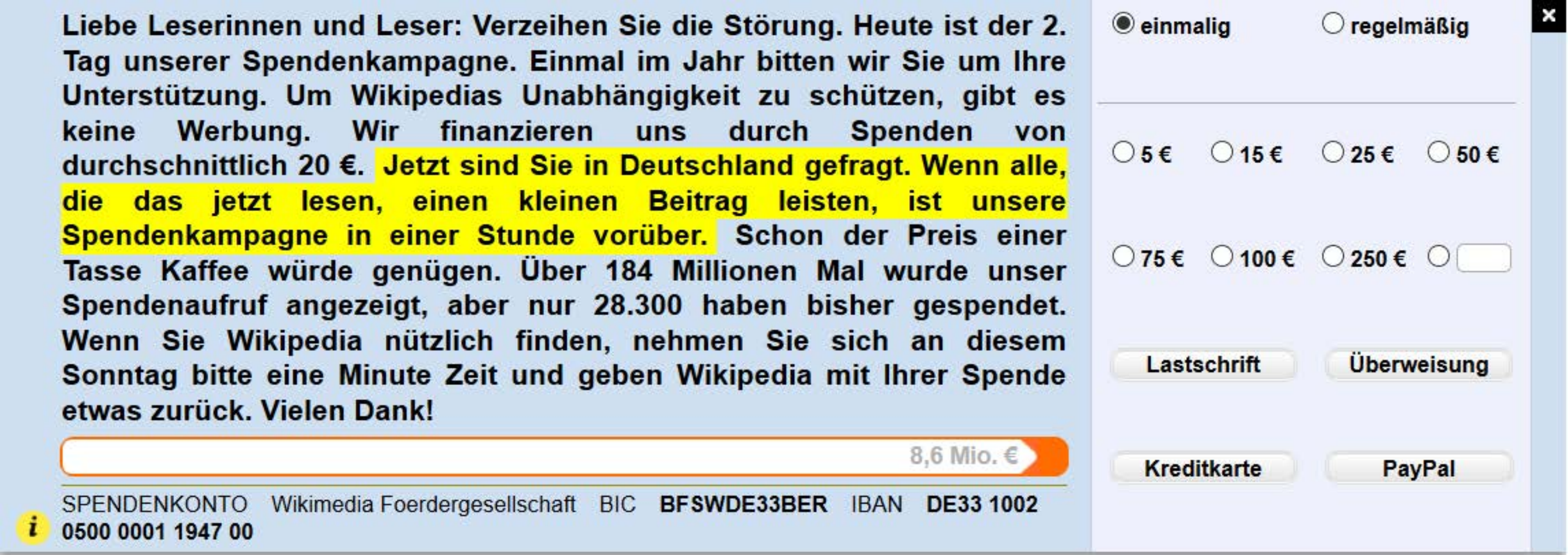}
\end{figure}

\begin{figure}[!ht]
\centering
\caption{Trial 3 -- Treatment}
\includegraphics[width=\linewidth]{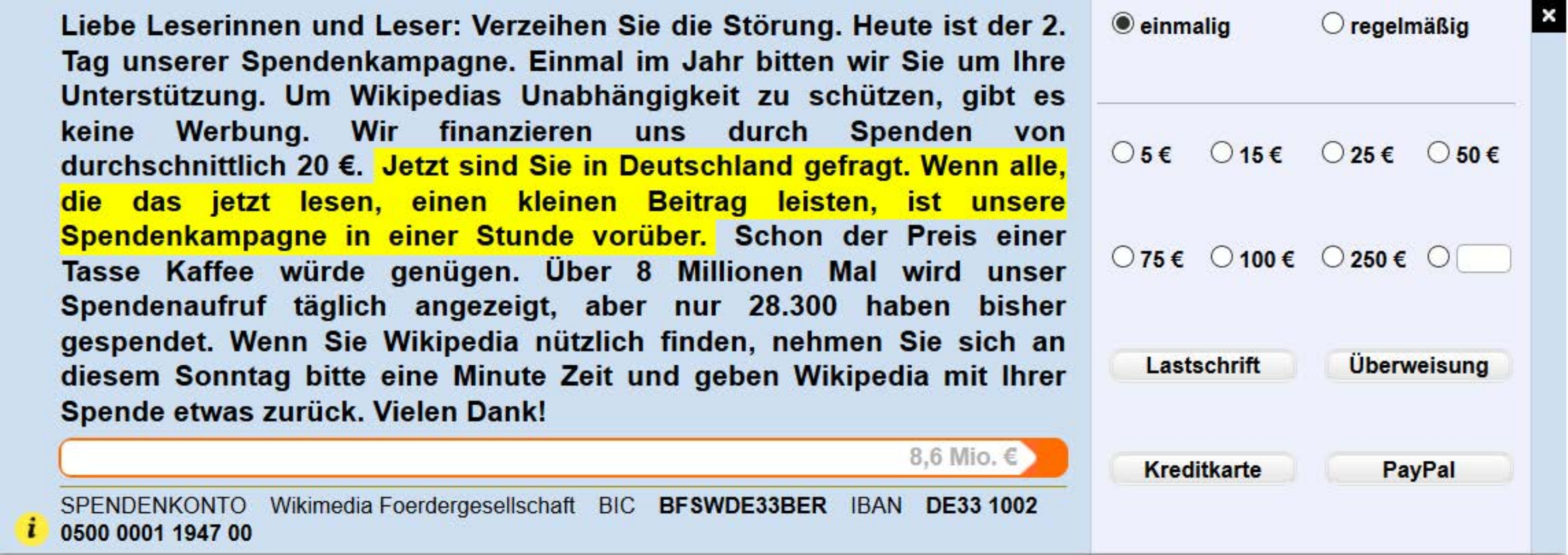}
\end{figure}

\clearpage

\begin{figure}[!ht]
\centering
\caption{Trial 4 -- Control}
\includegraphics[width=\linewidth]{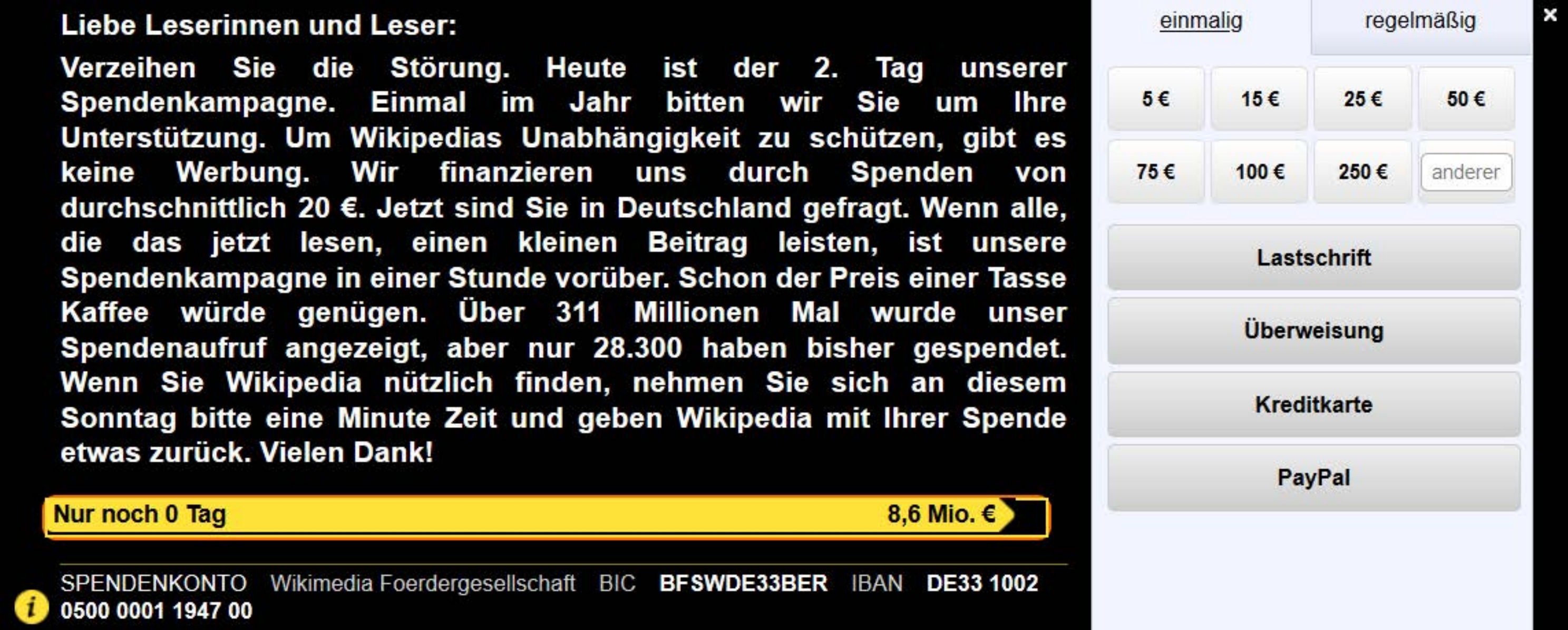}
\end{figure}

\begin{figure}[!ht]
\centering
\caption{Trial 4 -- Treatment}
\includegraphics[width=\linewidth]{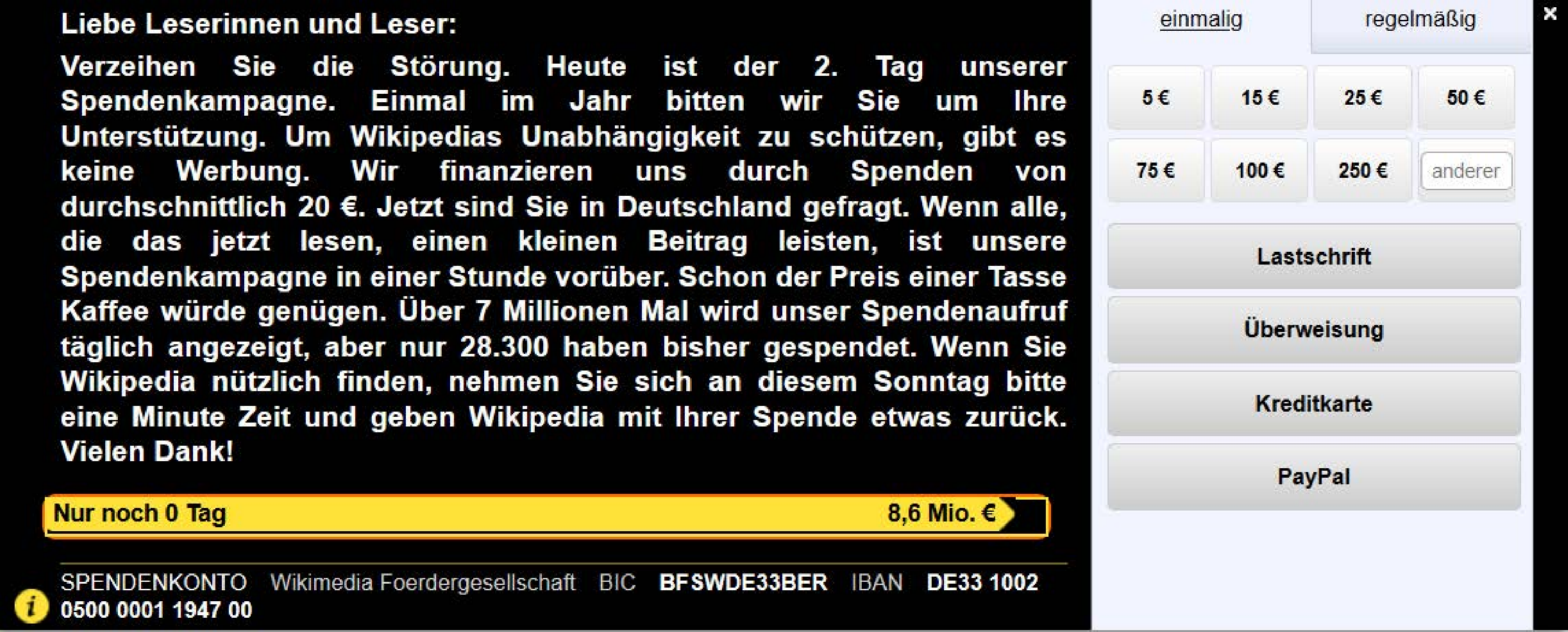}
\end{figure}

\clearpage

\begin{figure}[!ht]
\centering
\caption{Trial 5 -- Control}
\includegraphics[width=\linewidth]{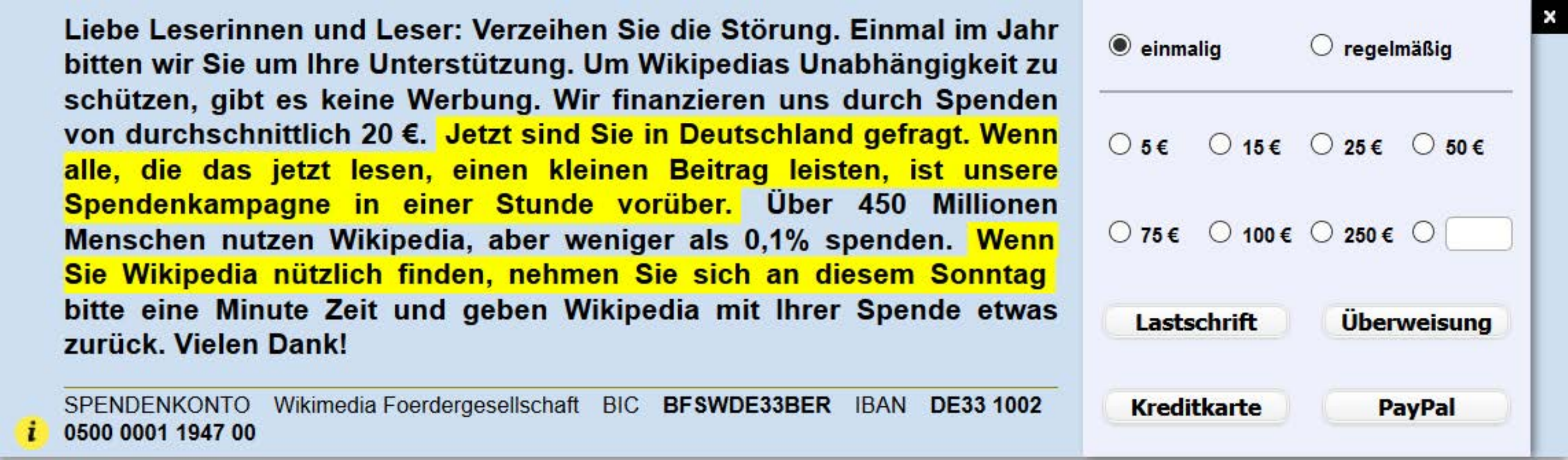}
\end{figure}

\begin{figure}[!ht]
\centering
\caption{Trial 5 -- Treatment}
\includegraphics[width=\linewidth]{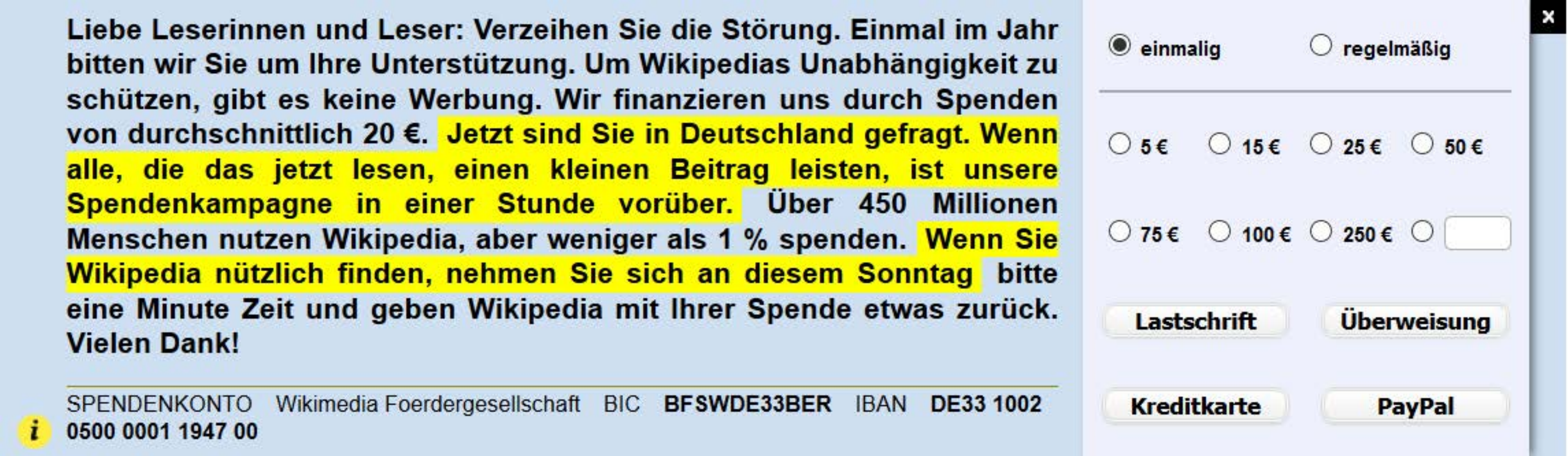}
\end{figure}

\clearpage

\begin{figure}[!ht]
\centering
\caption{Trial 6 -- Control}
\includegraphics[width=\linewidth]{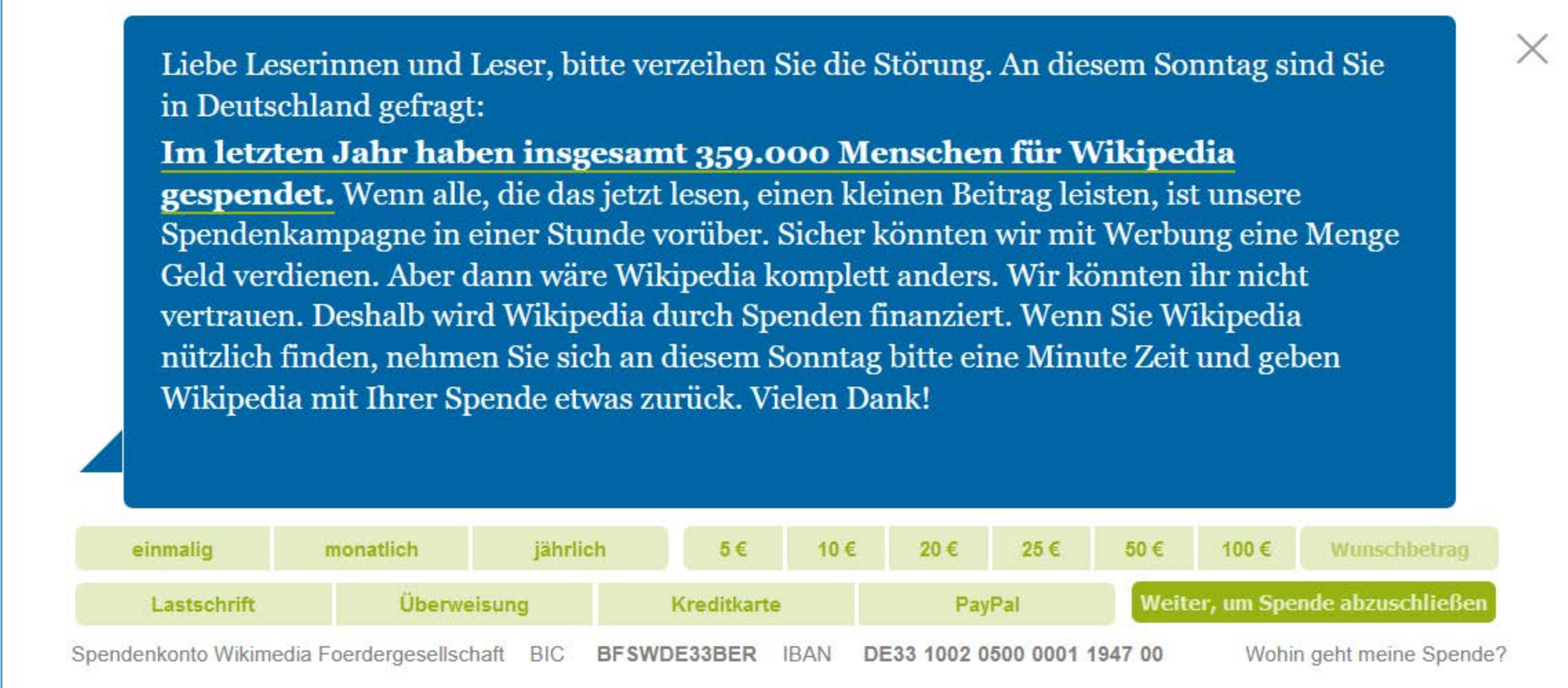}
\end{figure}

\begin{figure}[!ht]
\centering
\caption{Trial 6 -- Treatment}
\includegraphics[width=\linewidth]{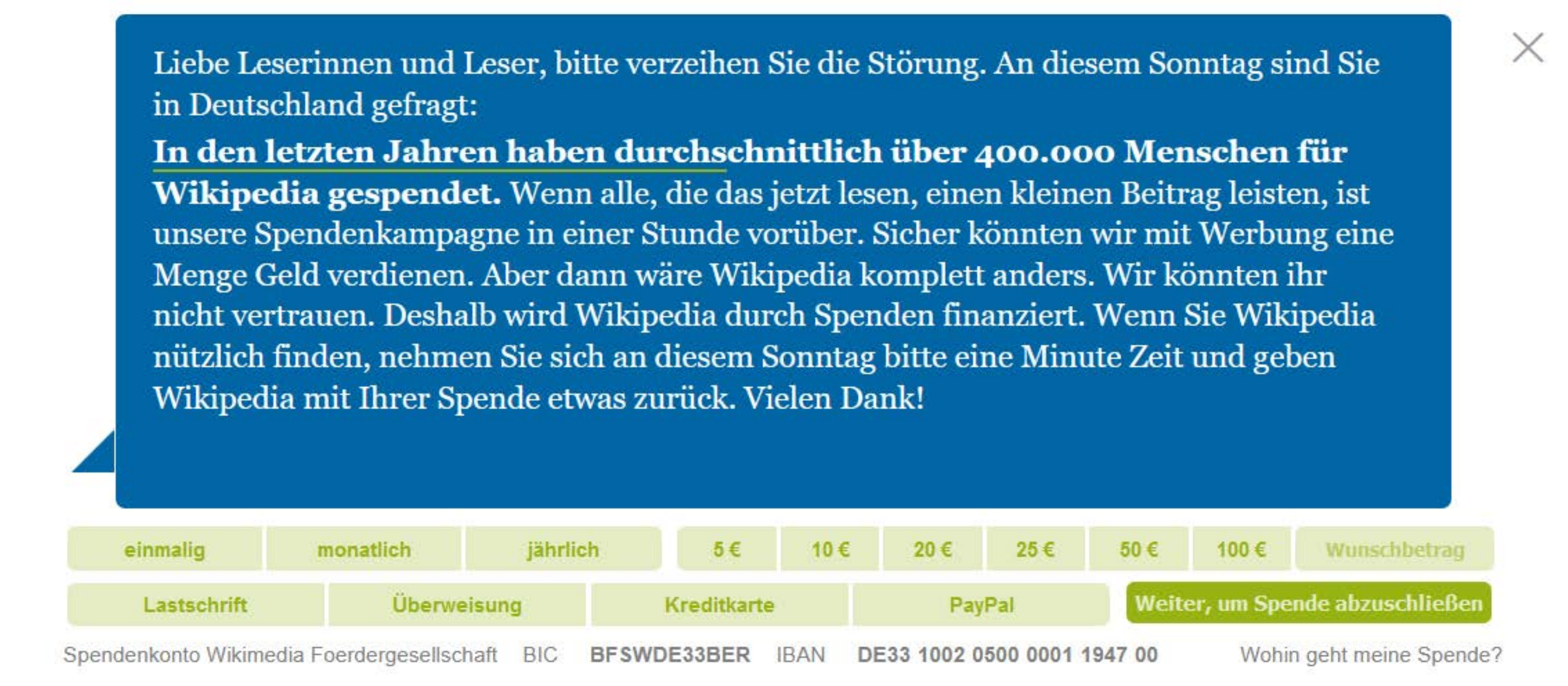}
\end{figure}

\emph{Translation:} \\
Dear Readers, please excuse the disturbance. This Friday we ask you in Germany: 
In the last year, a total of 359,000 people donated to Wikipedia. [\emph{Treatment banner}: Over the last years, more than 400,000 people donated to Wikipedia on average.] If everyone reading this gave a small amount, our campaign would end in an hour. Sure, we could make a lot of money with ads. But then Wikipedia wouldn't be the same. We wouldn't be able to trust it. This is why Wikipedia is funded by donations. If you consider Wikipedia useful, please take a minute and, with your donation, give something back to Wikipedia. Many thanks!

\emph{Note}: Given the German text of the treatment, it should implicitly be quite clear that the 400,000 refer to a \emph{yearly} average. However, this was not explicit in the text.

\end{document}

%% file: Wiki_paper1_Tab1_2021.tex
\begin{tabular}{@{}|c|l|p{0.55\textwidth}|c|c|c|c|c|@{}} \toprule\hline
\multicolumn{1}{|c|}{\textbf{Trial}} & \multicolumn{1}{|c|}{\textbf{Group}} & \multicolumn{1}{|c|}{\textbf{Treatment text}} & \multicolumn{1}{|c|}{\textbf{Number of}} & \multicolumn{1}{|c|}{\textbf{Number of}} & \multicolumn{1}{|c|}{\textbf{Total revenue}} & \multicolumn{1}{|c|}{\textbf{Average amount}} & \multicolumn{1}{|c|}{\textbf{Fundraising}} \\
& & \multicolumn{1}{|c|}{\emph{Click on the texts below to display}} & \multicolumn{1}{|c|}{\textbf{impressions}} & \multicolumn{1}{|c|}{\textbf{donations}} & \multicolumn{1}{|c|}{(in EUR)} & \multicolumn{1}{|c|}{\textbf{donated}} & \multicolumn{1}{|c|}{\textbf{Year}} \\
& & \multicolumn{1}{|c|}{\emph{the full banner text online.}} & & & & \multicolumn{1}{|c|}{(in EUR)} & \\ \hline\hline
& \multicolumn{7}{|l|}{\textbf{Variation in framing} (``only \ldots '' vs. ``already \ldots '')} \\ \hline\hline
\multirow{4}{*}{\textbf{1}} & control & \href{https://de.wikipedia.org/?banner=B14WMDE_141203_ctrl&uselang=en&force=1}{Our donation banner is viewed more than 20 million times a day, but only 115.000 people have donated so far} & 1,189,600 & 1,329 & 27,012 & 20.32 & \multirow{4}{*}{2014} \\ \cline{2-7}
& treatment & \href{https://de.wikipedia.org/?banner=B14WMDE_141203_20miovar&uselang=en&force=1}{Our donation banner is viewed more than 20 million times a day. Already 115.000 people have donated so far} & 1,198,100 & 1,095 & 22,283 & 20.35 & \\ \hline
\multirow{4}{*}{\textbf{2}} & control & \href{https://de.wikipedia.org/?banner=B15WMDE_29_151210_ctrl}{Our donation banner is viewed more than 8 million times a day, but only 28.300 people have donated so far} & 847,970 & 988 & 21,196 & 21.45 & \multirow{4}{*}{2015} \\ \cline{2-7}
& treatment & \href{https://de.wikipedia.org/?banner=B15WMDE_29_151210_var}{Our donation banner is viewed more than 8 million times a day. Already 28.300 people have donated so far} & 843,450 & 851 & 17,223 & 20.24 & \\ \hline\hline
& \multicolumn{7}{|l|}{\textbf{Variation in baseline numbers} (cumulative vs. daily banner impressions)} \\ \hline\hline
\multirow{4}{*}{\textbf{3}} & control & \href{https://de.wikipedia.org/?banner=B15WMDE_21_151202_fullimp}{Our donation banners have been viewed more than 184 million times, but only 28.300 people have donated so far} & 3,268,600 & 3,383 & 69,072 & 20.42 & \multirow{4}{*}{2015} \\ \cline{2-7}
& treatment & \href{https://de.wikipedia.org/?banner=B15WMDE_21_151202_ctrl}{Our donation banner is viewed more than 8 million times a day, but only 28.300 people have donated so far} & 3,271,190 & 3,222 & 69,676 & 21.62 & \\ \hline
\multirow{4}{*}{\textbf{4}} & control & \href{https://de.wikipedia.org/?banner=B15WMDE_41_151221_allimps}{Our donation banners have been viewed more than 311 million times, but only 28.300 people have donated so far} & 1,782,720 & 2,903 & 47,900 & 16.50 & \multirow{4}{*}{2015} \\ \cline{2-7}
& treatment & \href{https://de.wikipedia.org/?banner=B15WMDE_41_151221_ctrl}{Our donation banner is viewed more than 7 million times a day, but only 28.300 people have donated so far} & 1,782,740 & 2,916 & 46,474 & 15.94 & \\ \hline\hline
& \multicolumn{7}{|l|}{\textbf{Direct variation in social information} (donation rate and number of donors)} \\ \hline\hline
\multirow{4}{*}{\textbf{5}} & control & \href{https://de.wikipedia.org/?banner=B15WMDE_01_151012_fewer}{More than 450 million people use Wikipedia, but less than 0.1\% donate} & 616,400 & 1,881 & 38,259 & 20.34 & \multirow{4}{*}{2015} \\ \cline{2-7}
& treatment & \href{https://de.wikipedia.org/?banner=B15WMDE_01_151012_ctrl}{More than 450 million people use Wikipedia, but less than 1\% donate} & 606,000 & 1,874 & 36,452 & 19.45 & \\ \hline
\multirow{4}{*}{\textbf{6}} & control & \href{https://de.wikipedia.org/?banner=B18WMDE_10_181108_ctrl}{In the last year, a total of 359.000 people donated to Wikipedia} & 1,010,430 & 1,210 & 23,357 & 19.30 & \multirow{4}{*}{2018} \\
& & & & & & & \\ \cline{2-7}
& treatment & \href{https://de.wikipedia.org/?banner=B18WMDE_10_181108_var}{Over the last years, more than 400.000 people donated to Wikipedia on average} & 1,016,130 & 1,178 & 23,263 & 19.75 & \\ \hline \bottomrule
\end{tabular}

%% file: Wiki_paper1_2021-06-15-revision_withappendix.bbl
\begin{thebibliography}{37}
\newcommand{\enquote}[1]{``#1''}
\expandafter\ifx\csname natexlab\endcsname\relax\def\natexlab#1{#1}\fi

\bibitem[\protect\citeauthoryear{Allcott}{Allcott}{2011}]{Allcott2011}
\textsc{Allcott, H.} (2011): \enquote{{Social Norms and Energy Conservation},}
  \emph{Journal of Public Economics}, 95, 1082--1095.

\bibitem[\protect\citeauthoryear{Alpizar, Carlsson, and
  Johansson-Stenman}{Alpizar et~al.}{2008}]{alpizar_anonymity_2008}
\textsc{Alpizar, F., F.~Carlsson, and O.~Johansson-Stenman} (2008):
  \enquote{Anonymity, {Reciprocity}, and {Conformity}: {Evidence} from
  {Voluntary} {Contributions} to a {National} {Park} in {Costa} {Rica},}
  \emph{Journal of Public Economics}, 92, 1047--1060.

\bibitem[\protect\citeauthoryear{Antinyan and Asatryan}{Antinyan and
  Asatryan}{2019}]{Zareh2019}
\textsc{Antinyan, A. and Z.~Asatryan} (2019): \enquote{{Nudging for Tax
  Compliance: A Meta-Analysis},} {ZEW Discussion Paper No.~19-055}.

\bibitem[\protect\citeauthoryear{Bartke, Friedl, Gelhaar, and Reh}{Bartke
  et~al.}{2017}]{Bartke2017}
\textsc{Bartke, S., A.~Friedl, F.~Gelhaar, and L.~Reh} (2017): \enquote{Social
  comparison nudges: Guessing the norm increases charitable giving,}
  \emph{Economics Letters}, 152, 73--75.

\bibitem[\protect\citeauthoryear{Bernheim}{Bernheim}{1994}]{bernheim_theory_1994}
\textsc{Bernheim, B.~D.} (1994): \enquote{A {Theory} of {Conformity},}
  \emph{Journal of Political Economy}, 102, 841--877.

\bibitem[\protect\citeauthoryear{Bicchieri}{Bicchieri}{2005}]{bicchieri_2005}
\textsc{Bicchieri, C.} (2005): \emph{{The Grammar of Society: The Nature and
  Dynamics of Social Norms}}, Cambridge University Press.

\bibitem[\protect\citeauthoryear{Bicchieri and Dimant}{Bicchieri and
  Dimant}{2019}]{bicchieri_nudging_2019}
\textsc{Bicchieri, C. and E.~Dimant} (2019): \enquote{Nudging with {Care}:
  {The} {Risks} and {Benefits} of {Social} {Information},} \emph{Public
  Choice}, 1--22.

\bibitem[\protect\citeauthoryear{Bicchieri, Dimant, and Sonderegger}{Bicchieri
  et~al.}{2020}]{Bicchieri_2020}
\textsc{Bicchieri, C., E.~Dimant, and S.~Sonderegger} (2020): \enquote{{It's
  Not a Lie If You Believe the Norm Does Not Apply: Conditional Norm-Following
  with Strategic Beliefs},} CESifo Working Paper No.~8059.

\bibitem[\protect\citeauthoryear{Bicchieri and Xiao}{Bicchieri and
  Xiao}{2009}]{bicchieri_right_2009}
\textsc{Bicchieri, C. and E.~Xiao} (2009): \enquote{Do the {Right} {Thing}:
  {But} {Only} if {Others} {Do} {So},} \emph{Journal of Behavioral Decision
  Making}, 22, 191--208.

\bibitem[\protect\citeauthoryear{Bénabou and Tirole}{Bénabou and
  Tirole}{2006}]{benabou_incentives_2006}
\textsc{Bénabou, R. and J.~Tirole} (2006): \enquote{Incentives and {Prosocial}
  {Behavior},} \emph{American Economic Review}, 96, 1652--1678.

\bibitem[\protect\citeauthoryear{Bénabou and Tirole}{Bénabou and
  Tirole}{2011}]{benabou_identity_2011}
---\hspace{-.1pt}---\hspace{-.1pt}--- (2011): \enquote{Identity, {Morals}, and
  {Taboos}: {Beliefs} as {Assets},} \emph{Quarterly Journal of Economics}, 126,
  805--855.

\bibitem[\protect\citeauthoryear{Cantoni, Yang, Yuchtman, and Zhang}{Cantoni
  et~al.}{2019}]{Cantoni_2019}
\textsc{Cantoni, D., D.~Y. Yang, N.~Yuchtman, and Y.~J. Zhang} (2019):
  \enquote{{Protests as Strategic Games: Experimental Evidence from Hong Kong's
  Antiauthoritarian Movement},} \emph{Quarterly Journal of Economics}, 134,
  1021--1077.

\bibitem[\protect\citeauthoryear{Capraro and Rand}{Capraro and
  Rand}{2018}]{Capraro-and-Rand-2018}
\textsc{Capraro, V. and D.~G. Rand} (2018): \enquote{{Do the Right Thing:
  Experimental evidence that preferences for moral behavior, rather than equity
  or efficiency per se, drive human prosociality},} \emph{Judgment and Decision
  Making}, 13, 99--111.

\bibitem[\protect\citeauthoryear{Chen, Harper, Konstan, and Li}{Chen
  et~al.}{2010}]{Chen-et-al-2010}
\textsc{Chen, Y., M.~Harper, J.~Konstan, and S.~X. Li} (2010): \enquote{{Social
  Comparisons and Contributions to Online Communities: A Field Experiment on
  MovieLens},} \emph{American Economic Review}, 100, 1358--1398.

\bibitem[\protect\citeauthoryear{Croson and Shang}{Croson and
  Shang}{2013}]{croson_limits_2013}
\textsc{Croson, R. and J.~Shang} (2013): \enquote{Limits of the {Effect} of
  {Social} {Information} on the {Voluntary} {Provision} of {Public} {Goods}:
  {Evidence} from {Field} {Experiments},} \emph{Economic Inquiry}, 51,
  473--477.

\bibitem[\protect\citeauthoryear{d'Adda, Capraro, and Tavoni}{d'Adda
  et~al.}{2017}]{DADDA2017}
\textsc{d'Adda, G., V.~Capraro, and M.~Tavoni} (2017): \enquote{{Push, don’t
  nudge: Behavioral spillovers and policy instruments},} \emph{Economics
  Letters}, 154, 92--95.

\bibitem[\protect\citeauthoryear{Dimant}{Dimant}{2019}]{Dimant2019}
\textsc{Dimant, E.} (2019): \enquote{{Contagion of pro- and anti-social
  behavior among peers and the role of social proximity},} \emph{Journal of
  Economic Psychology}, 73, 66--88.

\bibitem[\protect\citeauthoryear{Dimant, {van Kleef}, and Shalvi}{Dimant
  et~al.}{2020}]{Dimant-et-al-2020}
\textsc{Dimant, E., G.~A. {van Kleef}, and S.~Shalvi} (2020): \enquote{Requiem
  for a Nudge: Framing effects in nudging honesty,} \emph{Journal of Economic
  Behavior \& Organization}, 172, 247--266.

\bibitem[\protect\citeauthoryear{Dur, Fleming, van Garderen, and van Lent}{Dur
  et~al.}{2021}]{Dur2021}
\textsc{Dur, R., D.~Fleming, M.~van Garderen, and M.~van Lent} (2021):
  \enquote{{A Social Norm Nudge to Save More: A Field Experiment at a Retail
  Bank},} \emph{Journal of Public Economics}, forthcoming.

\bibitem[\protect\citeauthoryear{Fellner, Sausgruber, and Traxler}{Fellner
  et~al.}{2013}]{Fellner_2013}
\textsc{Fellner, G., R.~Sausgruber, and C.~Traxler} (2013): \enquote{{Testing
  Enforcement Strategies in the Field: Threat, Moral Appeal and Social
  Information},} \emph{Journal of the European Economic Association}, 11,
  634--660.

\bibitem[\protect\citeauthoryear{Frey and Meier}{Frey and
  Meier}{2004}]{frey_social_2004}
\textsc{Frey, B.~S. and S.~Meier} (2004): \enquote{Social {Comparisons} and
  {Pro}-{Social} {Behavior}: {Testing} "{Conditional} {Cooperation}" in a
  {Field} {Experiment},} \emph{American Economic Review}, 94, 1717--1722.

\bibitem[\protect\citeauthoryear{Gallus}{Gallus}{2017}]{Gallus2017}
\textsc{Gallus, J.} (2017): \enquote{{Fostering Public Good Contributions with
  Symbolic Awards: A Large-Scale Natural Field Experiment at Wikipedia},}
  \emph{Management Science}, 63, 3999--4015.

\bibitem[\protect\citeauthoryear{Göbel and Munzert}{Göbel and
  Munzert}{2018}]{Munzert2018}
\textsc{Göbel, S. and S.~Munzert} (2018): \enquote{Political Advertising on
  the Wikipedia Marketplace of Information,} \emph{Social Science Computer
  Review}, 36, 157--175.

\bibitem[\protect\citeauthoryear{Gächter}{Gächter}{2007}]{gachter_conditional_2007}
\textsc{Gächter, S.} (2007): \enquote{Conditional {Cooperation}: {Behavioral}
  {Regularities} from the {Lab} and the {Field} and {Their} {Policy}
  {Implications},} in \emph{Economics and {Psychology}: {A} {Promising} {New}
  {Cross}-{Disciplinary} {Field}}, ed. by B.~S. Frey and A.~Stutzer, Cambridge,
  MA: MIT Press, {CESifo} {Seminar} {Series}, 19--50.

\bibitem[\protect\citeauthoryear{Goette and Tripodi}{Goette and
  Tripodi}{2021}]{Goette_2019}
\textsc{Goette, L. and E.~Tripodi} (2021): \enquote{{Social Influence} in
  {Prosocial Behavior: Evidence} from a {Large-Scale Experiment},}
  \emph{Journal of the European Economic Association}, forthcoming.

\bibitem[\protect\citeauthoryear{Greenstein and Zhu}{Greenstein and
  Zhu}{2012}]{Greenstein2012}
\textsc{Greenstein, S. and F.~Zhu} (2012): \enquote{Is Wikipedia Biased?}
  \emph{American Economic Review}, 102, 343--48.

\bibitem[\protect\citeauthoryear{Hager, Hensel, Hermle, and Roth}{Hager
  et~al.}{2019}]{Hager-et-al-2019}
\textsc{Hager, A., L.~Hensel, J.~Hermle, and C.~Roth} (2019):
  \enquote{{Political Activists as Free-Riders: Evidence from a Natural Field
  Experiment},} {IZA Discussion Paper No.~12759}.

\bibitem[\protect\citeauthoryear{Hallsworth, List, Metcalfe, and
  Vlaev}{Hallsworth et~al.}{2017}]{Hallsworth2017}
\textsc{Hallsworth, M., J.~List, R.~Metcalfe, and I.~Vlaev} (2017):
  \enquote{{The Behavioralist as Tax Collector: Using Natural Field Experiments
  to Enhance Tax Compliance},} \emph{Journal of Public Economics}, 148, 14--31.

\bibitem[\protect\citeauthoryear{Heldt}{Heldt}{2005}]{heldt_conditional_2005}
\textsc{Heldt, T.} (2005): \enquote{Conditional {Cooperation} in the {Field}:
  {Cross}-{Country} {Skiers}' {Behavior} in {Sweden},} Unpublished Manuscript.

\bibitem[\protect\citeauthoryear{List, Shaikh, and Xu}{List
  et~al.}{2019}]{List2019}
\textsc{List, J., A.~Shaikh, and Y.~Xu} (2019): \enquote{{Multiple Hypothesis
  Testing in Experimental Economics},} \emph{Experimental Economics}, 22,
  773--793.

\bibitem[\protect\citeauthoryear{List}{List}{2011}]{List2011}
\textsc{List, J.~A.} (2011): \enquote{{The Market for Charitable Giving},}
  \emph{Journal of Economic Perspectives}, 25, 157--80.

\bibitem[\protect\citeauthoryear{Martin and Randal}{Martin and
  Randal}{2008}]{martin_how_2008}
\textsc{Martin, R. and J.~Randal} (2008): \enquote{How is {Donation}
  {Behaviour} {Affected} by the {Donations} of {Others}?} \emph{Journal of
  Economic Behavior \& Organization}, 67, 228--238.

\bibitem[\protect\citeauthoryear{Potters, Sefton, and Vesterlund}{Potters
  et~al.}{2007}]{potters_leading-by-example_2007}
\textsc{Potters, J., M.~Sefton, and L.~Vesterlund} (2007):
  \enquote{Leading-{By}-{Example} and {Signaling} in {Voluntary} {Contribution}
  {Games}: {An} {Experimental} {Study},} \emph{Economic Theory}, 33, 169--182.

\bibitem[\protect\citeauthoryear{Shang and Croson}{Shang and
  Croson}{2009}]{shang_field_2009}
\textsc{Shang, J. and R.~Croson} (2009): \enquote{A {Field} {Experiment} in
  {Charitable} {Contribution}: {The} {Impact} of {Social} {Information} on the
  {Voluntary} {Provision} of {Public} {Goods},} \emph{The Economic Journal},
  119, 1422--1439.

\bibitem[\protect\citeauthoryear{Sunstein}{Sunstein}{2017}]{Sunstein2017}
\textsc{Sunstein, C.~R.} (2017): \enquote{{Nudges That Fail},}
  \emph{Behavioural Public Policy}, 1, 4–25.

\bibitem[\protect\citeauthoryear{Traxler}{Traxler}{2010}]{Traxler2010}
\textsc{Traxler, C.} (2010): \enquote{{Social norms and conditional cooperative
  taxpayers},} \emph{{European Journal of Political Economy}}, 26, 89--103.

\bibitem[\protect\citeauthoryear{Vesterlund}{Vesterlund}{2003}]{vesterlund_informational_2003}
\textsc{Vesterlund, L.} (2003): \enquote{The {Informational} {Value} of
  {Sequential} {Fundraising},} \emph{Journal of Public Economics}, 87,
  627--657.

\end{thebibliography}
